\documentclass[a4paper,11pt]{article}
\pdfoutput=1 

\usepackage{jcappub} 
\usepackage{amssymb}
\usepackage{xcolor}
\usepackage[T1]{fontenc} 
\usepackage{amsmath}
\usepackage{amsfonts}
\usepackage{mathrsfs}
\usepackage[scr=rsfso,cal=zapfc,frak=euler,bb=ams]{mathalfa}
\usepackage{trimclip}
\usepackage{accents}

\DeclareUnicodeCharacter{2212}{-}

\title{Relativistic massive compact stars supported by decoupled matter: Implications for mass-radius bounds }

\author[a,1]{S. K. Maurya
	\note{Corresponding author},}
\author[b,c]{A. Errehymy,}
\author[d]{Ksh. Newton Singh,}
\author[e]{G. Mustafa,}
\author[f]{Saibal Ray}

\affiliation[a]{Department of Mathematical and Physical Sciences,
	College of Arts and Sciences, University of Nizwa, P.O. Box 33, Nizwa 616, Sultanate of Oman}
\affiliation[b]{Astrophysics Research Centre, School of Mathematics, Statistics and Computer Science, University of KwaZulu-Natal, Private Bag X54001, Durban 4000, South Africa}
\affiliation[c]{Center for Theoretical Physics, Khazar University, 41 Mehseti Str., Baku, AZ1096, Azerbaijan}
\affiliation[d]{Department of Physics, National Defence Academy, Khadakwasla, Pune 411023, India}
\affiliation[e]{Department of Physics, Zhejiang Normal University, Jinhua 321004, China} 

\affiliation[f]{Centre for Cosmology, Astrophysics and Space Science (CCASS), GLA University, Mathura 281406, Uttar Pradesh, India}

\emailAdd{sunil@unizwa.edu.om}
\emailAdd{abdelghani.errehymy@gmail.com}
\emailAdd{ntnphy@gmail.com}
\emailAdd{gmustafa3828@gmail.com}
\emailAdd{saibal.ray@gla.ac.in}

\abstract{The merger of binary neutron stars (BNSs) is a remarkable astrophysical event where all four fundamental forces interplay dynamically across multiple stages, producing a rich spectrum of multi-messenger signals. These observations present a significant multiphysics modeling challenge but also offer a unique opportunity to probe the nature of gravity and the strong nuclear interaction under extreme conditions. The landmark detection of GW170817 provided essential constraints on the properties of non-rotating neutron stars (NSs), including their maximum mass ($M_{max}$) and radius distribution, thereby informing the equation of state (EOS) of cold, dense nuclear matter. While the inspiral phase of such events has been extensively studied, the post-merger signal holds even greater potential to reveal the behavior of matter at supranuclear densities, particularly in scenarios involving a transition to deconfined quark matter. Motivated by the recent gravitational wave event GW190814 ($2.5$–$2.67\,M_\odot$), we revisit the modeling of high-mass compact stars to investigate their internal structure via a generalized polytropic EOS. This framework incorporates a modified energy density profile and is coupled with the Tolman–Oppenheimer–Volkoff (TOV) equations. We explore mass–radius ($M$–$R$) relationships within both general relativity (GR) and the minimal geometric deformation (MGD) approach. Specifically, we constrain the radii of four massive compact objects---PSR J1614–2230 ($1.97^{+0.04}_{-0.04}\,M_\odot$), PSR J0952–0607 ($2.35^{+0.17}_{-0.17}\,M_\odot$), GW190814 ($2.5$–$2.67\,M_\odot$), and GW200210 ($2.83^{+0.47}_{-0.42}\,M_\odot$)---and demonstrate that our theoretical $M$–$R$ curves are consistent with observational data. These findings provide meaningful constraints on the EOS and underscore the potential of alternative gravity models to accommodate ultra-massive compact stars within a physically consistent framework.} 

\begin{document}
	
	\maketitle
	
	\flushbottom
	
\section{Introduction }
The merger of binary neutron stars (BNSs) represents a fascinating cosmic event where all four fundamental forces interact dynamically at different phases. This intricate interplay manifests itself in a variety of astronomical signals captured through multiple observational techniques. This exceptional attribute creates a formidable multiphysics challenge for precise modeling, yet it also opens up the possibility of constraining gravity and the strong interactions within dense matter. The historic multi-messenger detection of the BNS merger GW170817 \citep{LIGOScientific:2017vwq, Drout:2017ijr, Cowperthwaite:2017dyu} yielded pivotal insights into the properties of isolated non-rotating neutron stars (NSs), including constraints on their maximum mass $M_{TOV}$ and radius distribution \citep{Margalit:2017dij, Annala:2017llu, Radice:2017lry, Rezzolla:2017aly, Most:2018hfd}. These essential parameters can be leveraged to refine our understanding of the equation of state (EOS) for cold nuclear matter. Although many insights are derived from the gravitational wave (GW) signal observed during the inspiral phase, the post-merger signal is poised to deliver even deeper revelations about EOS under extreme density conditions \citep{Stergioulas:2011gd, Bauswein:2011tp, Bernuzzi:2015rla, Rezzolla:2016nxn, Breschi:2021xrx}. This becomes especially significant when considering the potential onset of a phase transition to quark matter \citep{Most:2018eaw, Bauswein:2018bma, Blacker:2020nlq, Prakash:2021wpz, Ujevic:2022nkr}. Moreover, the maximum mass $M_{\text{max}}$ of NSs and their radii are intrinsically linked through the EOS for dense matter, which can be explored by observational techniques. Observations are vital in establishing key constraints on $M_{\text{max}}$, with profound implications for our understanding of both astronomical phenomena and fundamental physics. Crucially, $M_{\text{max}}$ is primarily dictated by the EOS at densities greater than three times the nuclear saturation density, roughly $n_{\text{sat}} \approx 0.16 \, \text{fm}^{-3}$ \cite{Gandolfi:2011xu}. This highlights its role as a critical probe into the characteristics of matter at extreme densities. Determining the maximum mass $M_{\text{max}}$ is essential for probing the phases of cold and dense matter within the strongly coupled framework of quantum chromodynamics (QCD). This understanding allows for the development of the pressure versus energy density relationship--commonly referred to as the EOS--for these extreme phases. Conversely, the radii of canonical NSs with masses around $1.4 \, M_\odot$ are primarily dictated by the EOS at densities below approximately three times the nuclear saturation density $n_{\text{sat}}$ \citep{Lattimer:2000nx}. This contrast emphasizes the critical role of $M_{\text{max}}$ in unraveling the complex behaviors of matter under extreme conditions and highlights how different density regimes influence the structural properties of NSs. The maximum mass $M_{\text{max}}$ also determines the minimum mass required for a stellar black hole (BH), estimated to be around $\mathcal{O}(M_\odot)$. This factor is crucial in predicting the eventual outcomes of core-collapse supernovae and binary BNS mergers. The formation of a BH in core-collapse supernovae is influenced by the amount of fallback matter and is intricately linked to the properties of the progenitor star and the dynamics of neutrino emission during the proto-neutron star phase. In the case of BNS mergers, the formation of BH is determined by a combination of factors, including the total spiraling mass, the mass ejected and the interaction of rotational and magnetohydrodynamics forces \citep{Margalit:2017dij, Shibata:2019ctb}.

With the successful detection of several NS mergers via GW radiation and the expectation of many more in the near future, we are poised to gain significantly enhanced constraints on $M_{\text{max}}$. As GW detectors enhance their high-frequency capabilities, the ability to detect post-merger radiation will significantly deepen our understanding of $M_{\text{max}}$. Current insights into $M_{\text{max}}$ are crucial for elucidating the characteristics of the recently observed mergers GW190425 and GW190814, both of which exhibit evidence of a mass component greater than 2 $M_\odot$. This raises the possibility that the component could be either a heavy NS or a light BH. The simultaneous detection of electromagnetic signals from future GW events--mirroring the findings from the BNS merger GW170817 \citep{LIGOScientific:2017vwq, LIGOScientific:2018hze, LIGOScientific:2018cki}--will significantly enhance our understanding of $M_{\text{max}}$ \citep{Margalit:2017dij, Shibata:2019ctb}.

From a theoretical perspective, $M_{\text{max}}$ is essential to define the minimum and maximum radii as a function of NS mass $M$. Therefore, in addition to the valuable insights gained from radio and X-ray binary pulsar observations, which have accurately measured several NS masses and confirmed a lower limit of $M_{\text{max}} \gtrsim 2 \, M_\odot$ \citep{Demorest:2010bx, Antoniadis:2013pzd, Fonseca:2016tux, NANOGrav:2017wvv, NANOGrav:2019jur, Romani:2021xmb}, simultaneous analysis of GW and X-ray data determining NS masses and radii offers crucial constraints. To date, the radii derived from X-ray observations -- including those of quiescent low-mass X-ray binaries (QLMXBs) \citep{Rybicki:2005id}, photospheric radius expansion bursts (PREs) \citep{Ozel:2008kb}, and pulse profiles from rotation-powered millisecond pulsars \citep{Bogdanov:2006zd} -- as well as the inaugural GW detection of the BNS merger GW170817 \citep{LIGOScientific:2017vwq, LIGOScientific:2018hze}, predominantly pertain to NSs with canonical masses near 1.4 $M_\odot$. In light of this, the NS Interior Composition ExploreR (NICER) mission \citep{Miller:2016kae} seeks to measure the radii of notably massive NSs such as PSR J1614-2230 (approximately 1.91 $M_\odot$ \citep{Demorest:2010bx, Fonseca:2016tux, NANOGrav:2017wvv}) and PSR J0740+6620 (about 2.14 $M_\odot$ \citep{NANOGrav:2019jur}), which is highly relevant. Furthermore, forthcoming radio observations with the Square Kilometer Array (SKA) telescope \cite{Watts:2014tja} are anticipated to explore binary pulsars, potentially revealing even more massive NSs than previously identified.

The fourth observing run (O4) of the Advanced LIGO (aLIGO), Advanced Virgo, and KAGRA observatories kicked off on May 24, 2023, and wrapped up its first segment, known as O4a, shortly afterward. Right from the start, aLIGO picked up a GW signal called GW230529, which had a signal-to-noise ratio of 11.6. This signal is believed to come from the merger of a compact binary star system, with one star weighing between 2.5 and 4.5 $M_\odot$ and the other between 1.2 and 2.0 $M_\odot$, based on a 90\% credible interval \citep{LIGOScientific:2024elc}. Interestingly, the analysis of GW230529 does not give us any clues about the tidal deformability of the secondary star, while the primary star seems to have a very low tidal deformability. Figuring out the exact components of GW230529 using just GW data is quite tricky. However, if we take into account current estimates for the maximum mass of NSs \citep{Fan:2023spm}, it seems likely that GW230529 is the result of a NS merging with a BH. Additionally, the extended inspiral signals related to GW230529 could help us better understand any deviations from GR and even the potential presence of charge in BHs, as explored in the parametrized post-Einsteinian (ppE) framework \citep{Yunes:2009ke, Yunes:2016jcc}. 

The unique high-density environment of NSs has long been seen as a fantastic setting for exploring the properties of nuclear bulk matter, particularly EOS. However, researchers face significant technical challenges both in theoretically deriving the EOS through first-principles calculations based on QCD and experimentally constraining the EOS using nuclear or astrophysical measurements. Right now, we can better understand the low-density region of EOSs, at least up to saturation density, by looking at experimentally determined ground state properties of finite nuclei, such as binding energy per nucleon, charge radii, and neutron skin thickness (see, for example, \citep{Lattimer:2023rpe, MUSES:2023hyz}). An effective way to build an EOS that reflects these properties is through the relativistic mean field (RMF) approach \citep{Johnson:1955zz, Duerr:1956zz, Walecka:1974qa, Boguta:1977xi, Todd-Rutel:2005yzo, Mueller:1996pm}. This method creates EOSs that are highly parametrized on the basis of the strengths of various meson interactions, which are particularly responsive to different baryon densities within the EOS. Furthermore, in constant-coupling RMF parameterizations, the resulting EOSs are covariant, which helps to ensure that causality is maintained when we extend them to the neutron star regime, as shown by \cite{Mueller:1996pm}. This makes the RMF approach especially effective for integrating known nuclear experimental constraints at densities below and around saturation density, as well as astrophysical observational constraints at higher densities. The limitations on high-density NS EOS come from heavy-ion collision experiments \citep{LeFevre:2015paj, Morfouace:2019jky} and the observable characteristics of NSs, such as their maximum mass \citep{Demorest:2010bx, Antoniadis:2013pzd, Fonseca:2016tux, NANOGrav:2019jur, Sharma:2005mf, Oertel:2016bki, Khadkikar:2021yrj, Solanki:2021vnl, Das:2023sss, Panotopoulos:2024jtn}, tidal deformability \citep{LIGOScientific:2017vwq, LIGOScientific:2018hze} and radius measurements. In particular, the NICER x-ray mission has provided valuable simultaneous measurements of mass and radius for certain pulsars \citep{Miller:2019cac, Riley:2019yda, Miller:2021qha, Choudhury:2024xbk, Vinciguerra:2023qxq, Riley:2021pdl}, which have placed strong constraints on the EOS \citep{Raaijmakers:2019qny,Raaijmakers:2021uju, Rutherford:2024srk}. When looking at all these measurements together, a noticeable tension arises. The masses of neutron stars reach at least 2 $M_{\odot}$ \citep{Demorest:2010bx, Antoniadis:2013pzd, Fonseca:2016tux, NANOGrav:2019jur}, which suggests the need for a stiff EOS. In contrast, the tidal deformability observed for GW170817 points to a softer EOS \citep{LIGOScientific:2018hze}. However, EOSs that might initially seem too rigid to fit the tidal deformability constraint could actually align with it if a quark-hadron phase transition is taken into account \citep{Paschalidis:2017qmb, Montana:2018bkb, Christian:2018jyd, Sieniawska:2018zzj, Christian:2019qer}. This transition can create hybrid stars, which have an inner quark core surrounded by a nucleonic outer core and crust \citep{Ivanenko:1965dg, Itoh:1970uw, Schertler:2000xq, Alford:2004pf, Chen:2011my, Masuda:2012kf, Zacchi:2015oma}. These hybrid stars tend to show smaller tidal deformability values, making them more consistent with the characteristics of softer EOSs.

In Einstein gravity, the non-linear nature of the field equations can make finding exact solutions quite difficult. However, a promising method called the minimal geometric deformation (MGD) approach opens up new possibilities. This technique was developed within the framework of the Randall-Sundrum brane-world theory \citep{Ovalle:2013xla} and has been applied to various aspects of GR, including BHs, brane-world stars \citep{Ovalle:2014uwa, Casadio:2015jva}, gravitational lensing, and other stellar models. Then Ovalle~\cite{Ovalle:2017fgl} introduce gravitational decoupling approach so-called MGD to starts with a simple spherically symmetric source that has a known solution. This can then be used to couple with more complex sources, leading to an auxiliary system of equations to solve. By combining all independent solutions, the entire system can be effectively addressed. In practice, a function $f(r)$ is added to the radial metric potential, and the goal is to determine this unknown function. A key requirement for this method is that there should be no exchange of energy-momentum between the sources. An extension of the MGD approach~\cite{Ovalle:2018gic} allows for energy exchanges, making it even more versatile. Additionally, the MGD technique can be applied to anisotropic matter distributions. While isotropic stars have been the focus of many studies as equilibrium states in stellar evolution, models that consider different tangential and radial stresses are also important for understanding realistic scenarios. Anisotropy can arise from various factors, such as viscosity and strong magnetic fields, and can be incorporated into the system using the MGD method. Numerous researchers have explored this technique in different contexts. Significant work has also been done on solutions for the Yang-Mills-Einstein-Dirac field equations, axial symmetry, and hairy BHs using the MGD framework. 

Based on the above background, the motivation of the present investigation involved understanding the way matter distribution within the stars and how the physical process can be described by using a generalized polytropic EOS. Therefore, we adopt here the following outline: in Section (\ref{sec2}), we provide Einstein's field equation along with a brief study of the decoupling sector, and in Section (\ref{sec3}) we provide a minimally deformed solution for strange stars with a mimicking of the pressure constraint: $P_r(r)=\theta^1_1(r)$ (Subsection \ref{solA}), a mimicking of the density constraint: $\rho(r)=\theta^0_0(r)$ (Subsection \ref{solB}). Boundary conditions for strange star models under gravitational decoupling are applied in the physical configuration (\ref{sec4}). In Section (\ref{sec5}), a physical analysis of minimally deformed solutions is performed in strange star models, where physical behavior of energy density and pressure profiles (Subsection \ref{sec5.1}) and physical behavior of anisotropic profiles (Subsection \ref{sec5.2}), respectively, are discussed. We have done stability analysis in Section (\ref{sec6}), especially stability analysis via adiabatic index (Subsection \ref{sec6.1}), stability analysis via cracking criterion (Subsection \ref{sec6.2}) and stability analysis via Harrison-Zel'dovich-Novikov criteria (Subsection \ref{sec6.3}). The physical behavior of $M-R$ profiles is presented in Section (\ref{sec7}). A few concluding remarks are made in Section (\ref{sec8}).

\section{Einstein's field equation with decoupling sector: A brief study}\label{sec2}

We consider Einstein's field equation with the coupled unknown matter sector $\theta_{ij}$ as
\begin{eqnarray} \label{eq1}
 &&   G_{ij} \equiv R_{ij}-\frac{1}{2}\,g_{ij}\,\mathcal{R}= -8\pi (T_{ij}+\sigma\,\Theta_{ij}).
\end{eqnarray}

In this context, the relativistic units are represented as $G = c = 1$. The Ricci tensor is indicated by $R_{ij}$, where $\mathcal{R}$ signifies the contracted Ricci scalars and $\sigma$ indicates the decoupled parameter. It is possible to create scalar, vector, and tensor fields with the source $\theta_{ij}$, and the energy-momentum tensor is represented as $T_{ij}$. The satisfaction of the Bianchi identity by the Einstein tensor $(G_{ij})$ necessitates the conservation of $T_{ij}$ and $\Theta_{ij}$. Then we have
\begin{eqnarray}\label{eq3}
&& {\nabla_i}[\,T^{ij} + \sigma \Theta^{ij}]=0.
\end{eqnarray}

The interior region of the star system is described by the space-time of a particular static spherically symmetric line element:
\begin{eqnarray} \label{eq4}
 ds^2 = -e^ {\Omega (r)} dr^2 - r^2\big(d\theta^2 + \sin^2 \theta~ d\phi^2\big)~+ e^{\Psi( r)} dt^2,
 \end{eqnarray}
where the symbols $\Psi$ and $\Omega$ are functions of only $r$, known as metric potentials. 

In this regard, the distribution of matter $T_{ij}$ describes the anisotropic matter that is characterized by the effective energy-momentum tensor $T^{\text{eff}}_{ij}$, describes an anisotropic matter distribution.
\begin{eqnarray}
&& T^{\text{eff}}_{ij} = \left({\rho^{\text{eff}}}+{P^{\text{eff}}_t}\right)u_{i}u_{j}-{P^{\text{eff}}_t} g_{ij}+({P^{\text{eff}}_r}-{P^{\text{eff}}_t})\chi_{i}\,\chi_{j},\label{eq5}
\end{eqnarray}
where the formula for $u^i$ is defined as: $u^i=e^{\Psi(r)/2}{\delta^i}_4$, which denotes the four-velocity. The unit vector in the radial direction is denoted by the symbol $\chi^i$, which is defined as $\chi^i = e^{\Omega(r)/2} {\delta^i}_1$.

The pressures in the radial and tangential directions are represented by $P^{\text{eff}}_r$ and $P^{\text{eff}}_t$, respectively, while the energy density of matter is characterized by $\rho^{\text{eff}}$. Furthermore, the $4$-velocity $\chi^{i}$ and the unit space-like vector $u^{i}$ in the radial direction satisfied the conditions: $\chi^{}$.  Therefore, the components of the effective energy-momentum tensor $T^{\text{eff}}_{ij}$ can be provided as follows:
\begin{eqnarray}\label{eq6}
[T^0_0]^{\text{eff}}=\rho^{\text{eff}}, ~~[T^1_1]^{\text{eff}}=-P^{\text{eff}}_{r},~~\text{and}~~[T^2_2]^{\text{eff}}=-P^{\text{eff}}_t.~~~
\end{eqnarray}

Next, the complete formulation of Einstein's field equations can be expressed as the collection of the following differential equations for the metric~(\ref{eq1}):
\begin{eqnarray}
-8\pi [T^1_1]^{\text{eff}}&=& -\frac{1}{r^2}+e^{-\Omega}\left(\frac{1}{r^2}+\frac{\Psi^{\prime}}{r}\right)=8\pi\, P^{\text{eff}}_r,\label{eq9}\\
 -8\pi [T^2_2]^{\text{eff}}= -8\pi [T^3_3]^{\text{eff}} &=& \frac{e^{-\Omega}}{4}\left(2\Psi^{\prime\prime}+\Psi^{\prime2}-\Omega^{\prime}\Psi^{\prime}+2\frac{\Psi^{\prime}-\Omega^{\prime}}{r}\right)= 8\pi\, P^{\text{eff}}_t,~~\label{eq10} \\
8\pi [T^4_4]^{\text{eff}} &=&  \frac{1}{r^2}-e^{-\Omega}\left(\frac{1}{r^2}-\frac{\Omega^{\prime}}{r}\right)= 8\pi \rho^{\text{eff}}.\label{eq11}
\end{eqnarray}

Applying the metric function allows for the determination of the mass function, $m(r)$, of a charged perfect fluid sphere
\begin{equation}
e^{-\Omega} = 1 - \frac{2m(r)}{r},\label{eq12}
\end{equation}
one can also obtain as equivalent to~\citep{Sharp-Misner:1964} 
\begin{equation}
m(r)= \frac{1}{2} \int \rho^{\text{eff}}\, r^2 dr.\label{eq14}
\end{equation}

By using Eqs. (\ref{eq9}) and (\ref{eq12}), we derive
\begin{equation}
\frac{\Psi^\prime }{2}= \frac{8\pi r P^{\text{eff}}_r + 2m/r^2 }{1 - 2m/r }.\label{eq15}
\end{equation}

On the assumption that $P^{\text{eff}}_r \neq P^{\text{eff}}_t$, the condition $P^{\text{eff}}_r = P^{\text{eff}}_t$ suggests the presence of an isotropic fluid distribution. The anisotropy factor, represented by the symbol $\Delta$, is mathematically defined precisely as: $\Delta^{\text{eff}}=P^{\text{eff}}_t-P^{\text{eff}}_r$. The term $2(P^{\text{eff}}_t-P^{\text{eff}}_r)/r$ denotes a force that arises from the anisotropic properties of the fluid. When pressure $P^{\text{eff}}_t$ exceeds pressure $P^{\text{eff}}_r$, the force is directed in an outward direction. When the value of $P^{\text{eff}}_t$ is less than $P^{\text{eff}}_r$, the force acts in an inward direction. However, if the value of $P^{\text{eff}}_t$ is greater than the value of $P^{\text{eff}}_r$, the force allows the formation of a more condensed structure in the scenario of an anisotropic fluid as opposed to the isotropic fluid distribution~\citep{Gokhroo1994}. 

In addition to Eqs. (\ref{eq12}) and (\ref{eq15}), the pressure gradient may also be represented in terms of $m$, $q$, $\rho$, and $P_r$ using Eqs. (\ref{eq9})--(\ref{eq11})
\begin{equation}
\frac{dP^{\text{eff}}_r}{dr} =- \frac{8\pi r P^{\text{eff}}_r + 2m/r^2}{1 - 2m/r } (P^{\text{eff}}_r+\rho^{\text{eff}})+ \frac{2 (P^{\text{eff}}_t-P^{\text{eff}}_r)}{r},\label{eq7}
\end{equation}
which provides the generalized hydrostatic TOV equation for anisotropic stellar structure~\citep{TOV1,TOV2}.  

Our next approach is to determine a precise solution for the field equations (\ref{eq9})--(\ref{eq11}) for the TOV equation (\ref{eq7}) which describes a model of a strange star. We see that the system of field equations exhibits a significant degree of non-linearity, thereby presenting a challenging task in their solution.  Hence, we use an alternative method called gravitational decoupling using the minimum geometric deformation (MGD) methodology and a specific transformation associated with the gravitational potential
\begin{eqnarray}
&& \Psi(r) \longrightarrow \Phi(r)+\sigma\, \mathcal{G}(r), \label{eq26}\\
&& e^{-\Omega(r)} \longrightarrow \eta(r)+\sigma\, \mathcal{D}(r).  \label{eq27}
\end{eqnarray}
Let $\mathcal{G}(r)$ and $\mathcal{D}(r)$ represent the decoupling functions, respectively, in relation to the temporal and radial metric components. The deformation may be tuned appropriately by adjusting the decoupling constant $\sigma$. When $\sigma =0$, the conventional gravity theory GR is systematically restored. By using the MGD technique, we can establish the values of $\mathcal{G}(r)=0$ and $\mathcal{D}(r)\ne0$. This observation suggests that the appropriate transformation only applies to the radial component of the metric function, while the temporal component remains unaltered. This method of MGD partitions the decoupled system (\ref{eq9})--(\ref{eq11}) into two distinct parts. The initial system is associated with $T_{ij}$, while the second system is related to the additional source $\theta_{ij}$. To formulate the initial system, we examine the energy-momentum tensor $T_{ij}$ which characterizes an anisotropic matter distribution presented by
\begin{equation}\label{eq28}
T_{ij}=\left(\rho+P_t\right)u_{i}\,u_{j}-P_t\,\delta_{ij}+\left(P_r-P_t\right)\chi_{i}\,\chi_{j}.
\end{equation}

The energy density is indicated by $\rho$, while the radial pressure and the tangential pressure for the seed solution are denoted by $p_{r}$ and $p_{t}$. Accordingly, the effective quantities may be expressed as
\begin{eqnarray}
\rho^{\text{eff}}=\rho+\sigma \,\theta^0_0,~~P^{\text{eff}}_r=P_r-\sigma\,\theta^1_1,~~P^{\text{eff}}_t=P_t-\sigma\,\theta^2_2.~~~~  \label{eq29}
\end{eqnarray}

Furthermore, the related effective anisotropy is
\begin{eqnarray} 
&&\hspace{-0.7cm} \Delta^{\text{eff}}=P^{\text{eff}}_t-P^{\text{eff}}_r= \Delta_{GR}+\Delta_{\theta},
 \label{eq30}\\
&&\hspace{-0.7cm} \text{where}~~~~\Delta_{GR}= P_t-P_r~~~~~\text{and}~~~~~\Delta_\theta= \sigma (\theta^1_1-\theta^2_2)\nonumber. 
\end{eqnarray}

The effective anisotropy can be defined as the combined value of two anisotropies in relation to the matter distribution, namely $T_{ij}$ and $\theta_{ij}$.  Gravitational decoupling generates anisotropy ($\Delta_{\theta}$), which may increase the effective anisotropy. The system (\ref{eq9})--(\ref{eq11}) can be decomposed into two systems by employing the transformations (\ref{eq26}) and (\ref{eq27}). The first system is dependent on the gravitational potentials $\Phi$ and $\eta$, (viz. when $\beta = 0$):
\begin{eqnarray}
&&\hspace{-0.8cm}\rho= \frac{1}{8\pi} \bigg(\frac{1 }{r^2}-\frac{\eta  }{r^2}-\frac{\eta^{\prime} }{r} \bigg),\label{eq19}\\
&&\hspace{-0.8cm} P_r=\frac{1}{8\pi} \bigg(-\frac{1 }{r^2}+\frac{\eta  }{r^2}+\frac{\Phi^{\prime} \eta  }{r}\bigg), \label{eq20}\\
&&\hspace{-0.8cm} P_t=\frac{1}{8\pi} \bigg(\frac{\eta^{\prime} \Phi^{\prime}  }{4}+\frac{\Phi^{\prime \prime} \eta  }{2}+\frac{\Phi^{\prime 2} \eta }{4}  +\frac{\eta^{\prime}  }{2 r}+\frac{\Phi^{\prime} \eta  }{2 r}\bigg). \label{eq21}
\end{eqnarray}

From Eq. (\ref{eq7}), the following result is obtained:
\begin{eqnarray}
-\frac{\Phi^\prime}{2}(\rho+P_r)-P_r^{\prime}+\frac{2}{r}( P_{t}-P_r)=0.~~\label{eq22}
\end{eqnarray}

This is a TOV equation for the system's configuration (\ref{eq19})--(\ref{eq21}), a solution to which may be found in the spacetime presented below:
\begin{equation}\label{eq35}
ds^2=e^{\Phi(r)}dt^2-\frac{dr^2}{\eta(r)}-r^2d\theta^2+r^2\text{sin}^2\theta \,d\phi^2.
\end{equation}

By activating $\beta$, one may derive the second set of equations as
\begin{eqnarray}
&&\hspace{-0.7cm}\theta^{0}_0=-\frac{1}{8\pi}\Big(\frac{\mathcal{D}   }{r^2}+\frac{\mathcal{D}^\prime }{r}\Big), \label{eq36}\\
&&\hspace{-0.7cm}\theta^1_1=-\frac{1}{8\pi}\Big(\frac{\mathcal{D}  }{r^2}+\frac{\Phi^{\prime} \mathcal{D}   }{r}\Big), \label{eq37}\\
&&\hspace{-0.7cm}\theta^2_2=- \frac{1}{8\pi}\Big(\frac{1}{4} \mathcal{D}^\prime \Phi^{\prime}   +\frac{1}{2} \Phi^{\prime \prime} \mathcal{D}   +\frac{1}{4} \Phi^{\prime 2} \mathcal{D}  +\frac{\mathcal{D}^\prime  }{2 r}+\frac{\Phi^{\prime} \mathcal{D}  }{2 r}\Big). \label{eq38}
\end{eqnarray}

The following resulting equation is provided by the linear combination of the Eqs. (\ref{eq36})--(\ref{eq38}) as
\begin{eqnarray}
-\frac{\Phi^{\prime}}{2} (\theta^0_0-\theta^1_1)+ (\theta^1_1)^\prime+\frac{2}{r} ~(\theta^1_1-\theta^2_2)=0. \label{eq39}
\end{eqnarray}

The mass distribution for each system may be expressed by the following formula
\begin{eqnarray}
&&\hspace{-0.7cm} m_{Q}=\frac{1}{2} \int^r_0 \rho(x)\, x^2 dx~~~\text{and}~~~m_{\theta}= \frac{1}{2}\,\int_0^r \theta^0_0 (x)\, x^2 dx, \label{eq40}
\end{eqnarray}
where the mass functions corresponding to sources $T_{ij}$ and $\theta_{ij}$ are indicated by the variables $m_{GR}(r)$ and $m_{\theta}(r)$, respectively. 
 
Finally, the benefit of MGD-decoupling gets clear: we can generalize any existing solutions that are linked to the matter-spacetime $\{T_{ ij}, \Phi, \eta\}$ provided by Eqs. (\ref{eq9})--(\ref{eq11}) and by solving the unconventional gravitational system of the equations Eqs. (\ref{eq36})--(\ref{eq38}) to find $\{\theta_{ ij }$, $\mathcal{G}$, $\mathcal{D}\}$. Therefore, we may produce the "$\theta$-version" of any $\{T_{ij} \Phi, \eta\}$-solution as 
\begin{eqnarray}
\{T_{ ij},~ \Phi(r),~ \eta(r)\} \Longrightarrow \{T^{\text{eff}}_{ ij}, ~\Psi(r),~~\Omega(r)\}.
\end{eqnarray}

The above connection outlines a direct approach to investigate the effects of gravity that extend beyond standard Einstein gravity.

\section{Minimally deformed solution for compact star (CS) } \label{sec3}

In this Section, we will solve both systems of equations (\ref{eq36})--(\ref{eq39}) and (\ref{eq19})--(\ref{eq21}) related to the sources ${T}_{\mu\nu}$ and $\theta_{\mu\nu}$. The energy-momentum tensor ${T}_{ij}$ describes an anisotropic fluid matter distribution; therefore, $\theta_{ij}$ may enhance the total anisotropy of the system, which helps prevent the gravitational collapse of the system. Furthermore, if we look at the second system, it shows clearly that the solution of the second system depends on the first system.  Then it is mandatory to initially solve the first system. For solving the first system (\ref{eq36})--(\ref{eq39}), we use a generalized polytropic equation of state (EOS) of the form,
\begin{eqnarray}
P_r &=& \alpha\, \rho^{1+1/n}+\beta \rho + \gamma, \label{eq51}
\end{eqnarray}
where $\alpha,~ \beta$ and $\gamma$ are constant parameters with proper dimensions and $n$ denotes a polytropic index. Let us delve into the evolutionary background of the polytropic EOS, represented as $P_r=\alpha~\rho^{1+\frac{1}{n}}$. This formulation has been extensively used to analyze the physical characteristics of compact stellar objects in a variety of scenarios \citep{1964ApJ...140..434T, PhysRevD.4.2185, PhysRevD.88.084022, PhysRevD.93.024047, maurya2023effect, cosenza1981some, Azam:2017vyb}.

The component given by $P_r=\alpha~\rho^{1+\frac{1}{n}}$ could come from different origins. It might be related to Bose-Einstein condensates, which can exhibit repulsive interactions when $\alpha>0$ or attractive interactions when $\alpha<0$. Alternatively, there may be other sources that contribute to this formulation \citep{Chavanis:2012kla}. In cosmology, the barotropic EOS $P_r = \beta~\rho$ has several interpretations: $\beta=1$ indicates stiff fluid matter, $\beta = 1/3$ represents radiation, $\beta = 0$ corresponds to dust, $\beta = -2/3$ signifies quintessence, $\beta=-1$ denotes vacuum energy and $\beta = -4/3$ relates to the phantom field. The polytropic EOS (\ref{eq51}) can represent the MIT bag model with parameters $ \alpha = 0$, $ \beta = 1/4 $, and $ \gamma = -4\,\mathcal{B}_g/3 $, where $ \mathcal{B}_g $ is a constant of the bag \citep{Farhi:1984qu}. This emphasizes the role of $\alpha$ in the model. To address the complexity of exact solutions, we use a polytropic index of 1, resulting in a quadratic term $\alpha\rho^2$ that represents the neutron liquid as a Bose-Einstein condensate. The linear terms $ \beta \rho + \gamma $ come from the free-quarks model for $ \beta = 1/4 $ and $ \gamma = -4\,\mathcal{B}_g/3 $. Consequently, these NSs are likely \textit{hybrid stars}. In 2014, \cite{Chavanis:2012kla} looked into the polytropic EoS with negative indices in the late universe, linking it to intriguing ideas like quantum fluctuations and constant density cosmology \citep{Freitas:2013nxa}. This approach was also applied to explore self-bound compact objects \citep{1964ApJ...140..434T, Azam:2017vyb, Mardan:2020noh}.

In order to get a non-singular, monotonically decreasing matter density within the spherically symmetric star system, we choose a modified version of $\rho$ as proposed by Mak \& Harko~\cite{Harko:2002prx}:
 \begin{eqnarray}
     \rho(r)=\rho_0 \bigg[ 1-\bigg(1-\frac{\rho_s}{\rho_0}\bigg) \frac{r^2}{R^2} \bigg], \label{eq2.1}
 \end{eqnarray}
where the constants $\rho_0$ and $\rho_s$ represent the maximum and minimum values of the central and surface of $\rho$, respectively.

Now, by using Eqs. (\ref{eq39}) and (\ref{eq40}) in Eq. (\ref{eq51}), one can find the differential equation as follows:
\begin{eqnarray} 
-120 \pi  \alpha  \rho_0^2 r \left(r^2-R^2\right)^2+8 \pi  \rho_0 r \big[30 \alpha  \rho_s r^4+3 r^2 R^2 (5 \beta -10 \alpha  \rho_s+1)+3 r^3 R^2 \Phi'(r)-5 r R^4 \Phi'(r)\nonumber\\-5 (3 \beta +1) R^4\big]-3 \left(40 \pi  \alpha  \rho_s^2 r^5+8 \pi  \rho_s r^3 R^2 \left(5 \beta +r \Phi'(r)+1\right)-5 R^4 \left(\Phi'(r)-8 \pi  \gamma  r\right)\right)=0.~~~~~\label{eq52}
\end{eqnarray}

The Eq. (\ref{eq52}) leads directly to
\begin{equation} \label{EOS1}
\Phi(r) = \int{\mathcal{F}({\bar r)\,{ d\bar r}}},~~\end{equation}  
where
\begin{eqnarray} 
\label{EOS2}
\mathcal{F}(r)  &=& \frac{8} {24 \pi  \rho_0 r^4 R^2-40 \pi  \rho_0 r^2 R^4-24 \pi  \rho_s r^4 R^2+15 R^4}\Big[15 \pi  \alpha  \rho_0^2 r^5-30 \pi  \alpha  \rho_0^2 r^3 R^2+15 \pi  \alpha  \rho_0^2 r R^4\nonumber\\&&-30 \pi  \alpha  \rho_0 \rho_s r^5+30 \pi  \alpha  \rho_0 \rho_s r^3 R^2-15 \pi  \beta  \rho_0 r^3 R^2-3 \pi  \rho_0 r^3 R^2+15 \pi  \beta  \rho_0 r R^4+5 \pi  \rho_0 r R^4\nonumber\\&&+15 \pi  \alpha  \rho_s^2 r^5+15 \pi  \beta  \rho_s r^3 R^2+3 \pi  \rho_s r^3 R^2+15 \pi  \gamma  r R^4\Big]. \nonumber
\end{eqnarray}

On the other hand, matter density profile (\ref{eq2.1}) along with the Eq. (\ref{eq19}) provides the other potential $\eta(r)$, as
\begin{eqnarray}
    \eta (r)= \frac{8}{15} \pi  \rho_0 r^2 \left(\frac{3 r^2}{R^2}-5\right)-\frac{8 \pi  \rho_s r^4}{5 R^2}+1. 
\end{eqnarray}

After integration of the above differential equation (\ref{EOS1}), we find the potential $\Phi(r)$, in the following form:
\begin{eqnarray}
\Phi(r)&=&\frac{\sqrt{5}  \left(52 \pi  \alpha  \rho_0^2 R^2-45 \alpha  \rho_0+12 \pi  (\beta +1) \rho_0 R^2+9 \left(5 \alpha  \rho_s+8 \pi  \gamma  R^2\right)\right)}{12\sqrt{2 \pi } R \sqrt{\rho_0 \left(10 \pi  \rho_0 R^2-9\right)+9 \rho_s}}\nonumber\\&& \times \tanh ^{-1}\left(\frac{\sqrt{\frac{2 \pi }{5}} \left(6 r^2 (\rho_s-\rho_0)+5 \rho_0 R^2\right)}{R \sqrt{\rho_0 \left(10 \pi  \rho_0 R^2-9\right)+9 \rho_s}}\right)+\frac{5 \alpha  r^2 (\rho_0-\rho_s)}{2 R^2}-\frac{1}{12} (15 \beta +5 \alpha  \rho_0+3) \nonumber\\&& \times \log \left\{24 \pi  r^4 (\rho_0-\rho_s)+5 R^2 \left(3-8 \pi  \rho_0 r^2\right)\right\}+\mathcal{C},  \label{eq53} 
\end{eqnarray}
where $C$ is the constant of integration. Using the expressions for $W(r)$ and $H(r)$, we find the expressions for $\rho$, $P_r$, and $P_t$ as
\begin{eqnarray} \label{eq55}
&&\hspace{-0.15cm} P_r(r) = \frac{1}{R^4}\left[\gamma\,R^4+\left(r^2 (\rho_s-\rho_0)+\rho_0 R^2\right) \left(\alpha  r^2 (\rho_s-\rho_0)+\alpha  \rho_0 R^2+\beta  R^2\right)\right],\\ 
&&\hspace{-0.15cm} P_{_t}(r)=-\frac{1}{R^6 \left(-24 \pi  \rho_0 r^4+40 \pi  \rho_0 r^2 R^2+24 \pi  \rho_s r^4-15 R^2\right)}\Big[30 \pi  \alpha ^2 \rho_0^4 r^2   \left(r^2-R^2\right)^4-4 \pi  \alpha  \rho_0^3 r^2 \nonumber\\&&\hspace{0.8cm}\times\left(r^2-R^2\right) \big(30 \alpha  \rho_s r^6-3 r^4 R^2  (-5 \beta +20 \alpha  \rho_s+3)+r^2 R^4 (-30 \beta +30 \alpha  \rho_s+17)+15 \beta  R^6\big)\nonumber\\&&\hspace{0.8cm}+\rho_0^2 \big[2 \pi  r^2 \big(90 \alpha ^2 \rho_s^2 r^8+r^4 R^4 [30 \alpha  \gamma +15 \beta ^2-6 \beta  +90 \alpha ^2 \rho_s^2 -4 \alpha  (45 \beta -26) \rho_s+3]-18 \alpha  \rho_s r^6 \nonumber\\&&\hspace{0.8cm} \times R^2 (3-5 \beta +10 \alpha  \rho_s)-2 r^2 R^6 \big(30 \alpha  \gamma +15 \beta ^2-7 \beta +\alpha  (17-45 \beta ) \rho_s+4\big) +5 R^8 (6 \alpha  \gamma +3 \beta ^2\nonumber\\&&\hspace{0.8cm}+1)\big)+15 \alpha  R^4 \left(3 r^4-4 r^2 R^2+R^4\right)\big]+\rho_0 \Big[-120 \pi  \alpha ^2 \rho_s^3 r^8 \left(r^2-R^2\right)-4 \pi  \alpha  \rho_s^2 r^6 R^2 \big(9 (5 \beta -3) r^2\nonumber\\&&\hspace{0.8cm}+(26-45 \beta ) R^2\big)-2 \rho_s r^2 R^4  \big(6 \pi  r^4 \left(10 \alpha  \gamma +5 \beta ^2-2 \beta +1\right)-2 \pi  r^2 R^2 \left(30 \alpha  \gamma +15 \beta ^2-7 \beta +4\right)\nonumber\\&&\hspace{0.8cm}+15 \alpha  \big(3 r^2-2 R^2\big)\big) +3 R^6 \big\{-4 \pi  (5 \beta +1) \gamma  r^4+10 \beta  r^2 \left(2 \pi  \gamma  R^2-1\right)+5 \beta  R^2\big\}\Big]+3 \Big\{10 \pi  \alpha ^2 \rho_s^4 r^{10}\nonumber\\&&\hspace{0.8cm}+4 \pi  \alpha  (5 \beta -3) \rho_s^3 r^8 R^2+\rho_s^2 r^4 R^4  \left(15 \alpha +2 \pi  r^2 \left(10 \alpha  \gamma +5 \beta ^2-2 \beta +1\right)\right)+2 \rho_s r^2 R^6 \big[5 \beta (2 \pi  \gamma  r^2\nonumber\\&&\hspace{0.8cm}+1)+2 \pi  \gamma  r^2\big]+5 \gamma  R^8 \left(2 \pi  \gamma  r^2+1\right)\Big\}\Big]. \label{eq57}
\end{eqnarray}

\subsection{Mimicking of the Pressure Constraint: $P_r(r)=\theta^1_1(r)$} \label{solA}

By mimicking the seed pressure ($P_r$) to the component ($\theta^1_1$) through Eqs. (\ref{eq20}) and (\ref{eq37}), we get a directly exact solution for $\mathcal{D}(r)$ as
\begin{eqnarray}
  && \hspace{-0.3cm}  \mathcal{D}(r)=\frac{-8 \pi  r^2 \left\{8 \pi  \rho_0 r^2 \left(3 r^2-5 R^2\right)-24 \pi  \rho_s r^4+15 R^2\right\} }{15 R^2 \left[R^4 \mathcal{D}_1(r)+8 \pi  \alpha  r^6 (\rho_0-\rho_s)^2-8 \pi  r^4 R^2 (\rho_0-\rho_s) (\beta +2 \alpha  \rho_0)\right]}\nonumber\\&&\hspace{0.8cm} \times \Big[\alpha  \rho_0^2 \left(r^2-R^2\right)^2-\rho_0 \left(r^2-R^2\right) \left(2 \alpha  \rho_s r^2+\beta  R^2\right)+\alpha  \rho_s^2 r^4+\beta  \rho_s r^2 R^2+\gamma  R^4\Big].~~~~~~\label{eq2.18}
\end{eqnarray}
where, $\left(8 \pi  \alpha  \rho_0^2 r^2+8 \pi  \beta  \rho_0 r^2+8 \pi  \gamma  r^2+1\right)$.\\
For this pressure constraint case, the deformed MGD solution can be described by the following spacetime geometries
\begin{eqnarray}
 && \hspace{-0.3cm}  e^{\Phi(r)}= \exp\Bigg[\frac{\sqrt{5}  \left\{52 \pi  \alpha  \rho_0^2 R^2-45 \alpha  \rho_0+12 \pi  (\beta +1) \rho_0 R^2+9 \left(5 \alpha  \rho_s+8 \pi  \gamma  R^2\right)\right\}}{12\sqrt{2 \pi } R \sqrt{\rho_0 \left(10 \pi  \rho_0 R^2-9\right)+9 \rho_s}}\nonumber\\&&\hspace{1.3cm} \times \tanh ^{-1}\left(\frac{\sqrt{\frac{2 \pi }{5}} \left(6 r^2 (\rho_s-\rho_0)+5 \rho_0 R^2\right)}{R \sqrt{\rho_0 \left(10 \pi  \rho_0 R^2-9\right)+9 \rho_s}}\right)+\frac{5 \alpha  r^2 (\rho_0-\rho_s)}{2 R^2}-\frac{1}{12} (15 \beta +5 \alpha  \rho_0+3) \nonumber\\&&\hspace{1.3cm} \times \log \left(24 \pi  r^4 (\rho_0-\rho_s)+5 R^2 \left(3-8 \pi  \rho_0 r^2\right)\right)+\mathcal{C}\Bigg]=e^{\Psi(r)},\label{eq2.19}\\
  && \hspace{-0.3cm}    e^{-\Omega(r)}= \eta(r)+\sigma\,\mathbf{D}(r)= \bigg[\frac{8}{15} \pi  \rho_0 r^2 \left(\frac{3 r^2}{R^2}-5\right)-\frac{8 \pi  \rho_s r^4}{5 R^2}+1\bigg]+ \sigma \bigg[\Big(\alpha  \rho_0^2  \left(r^2-R^2\right)^2\nonumber\\&&\hspace{1.8cm}-\rho_0 \left(r^2-R^2\right) \left(2 \alpha  \rho_s r^2+\beta  R^2\right)+\alpha  \rho_s^2 r^4+\beta  \rho_s r^2 R^2+\gamma  R^4\Big) \nonumber\\&&\hspace{1.8cm} \times \frac{-8 \pi  r^2 \left(8 \pi  \rho_0 r^2 \left(3 r^2-5 R^2\right)-24 \pi  \rho_s r^4+15 R^2\right) }{15 R^2 \left[R^4 \mathcal{D}_1(r)+8 \pi  \alpha  r^6 (\rho_0-\rho_s)^2-8 \pi  r^4 R^2 (\rho_0-\rho_s) (\beta +2 \alpha  \rho_0)\right]}\bigg] .~~~~~~ \label{eq2.20}
\end{eqnarray}

The expressions for the components of the new source $\theta_{ij}$ as:
\begin{small}
\begin{eqnarray}
 &&\hspace{-0.15cm}   \theta^0_0(r)=\frac{1}{15 \mathcal{F}_1(r)}\Big(-24 \rho_s \pi  r^4+8 \rho_0 \pi  \left(3 r^2-5 R^2\right) r^2+15 R^2\Big) \Big(\rho_s^2 \alpha  r^4+\rho_s R^2 \beta  r^2+\rho_0^2 \nonumber\\&&\hspace{0.8cm} \times \left(r^2-R^2\right)^2 \alpha  -\rho_0 \left(r^2-R^2\right)  \left(2 \rho_s \alpha  r^2+R^2 \beta \right)+R^4 \gamma \Big)-\frac{1}{15\mathcal{F}_2(r)} \mathcal{G}_{0}(r), \label{eq2.21}\\
  &&\hspace{-0.15cm}   \theta^1_1(r)=\frac{1}{R^4}\left[\alpha  \rho_0^2 \left(r^2-R^2\right)^2-\rho_0 \left(r^2-R^2\right) \left(2 \alpha  \rho_s r^2+\beta  R^2\right)+\alpha  \rho_s^2 r^4+\beta  \rho_s r^2 R^2+\gamma  R^4\right],~~~~~~\label{eq2.22}\\
     &&\hspace{-0.15cm} \theta^2_2(r)= -\frac{1}{15 R^6 \left(-24 \rho_0 \pi  r^4+24 \rho_s \pi  r^4+40 \rho_0 \pi  R^2 r^2-15 R^2\right) \mathcal{F}_3(r)}\bigg[28800 \rho_0^8 \pi ^3 r^6 \alpha ^4 \left(r^2-R^2\right)^8\nonumber\\&&\hspace{0.8cm}-3840 \rho_0^7 \pi ^3 r^6 \alpha ^3\big(60 \rho_s \alpha  r^6-3 R^2 (40 \rho_s \alpha -10 \beta +3) r^4+R^4 (60 \rho_s \alpha -60 \beta +17) r^2+30 R^6 \beta \big) \nonumber\\&&\hspace{0.8cm} \times \left(r^2-R^2\right)^5-240 \rho_0^5 \pi ^2 r^4 \alpha  \mathcal{F}_5(r) \left(r^2-R^2\right)^3+4 \rho_0^4 \pi  r^2  \mathcal{F}_6(r) \left(r^2-R^2\right)+240 \rho_0^6 \pi ^2 \left(r^3-r R^2\right)^4 \nonumber\\&&\hspace{0.8cm} \times \alpha ^2 \left(15 \left(13 r^4-18 R^2 r^2+5 R^4\right) \alpha  R^4+8 \pi  r^2 \mathcal{F}_7(r)\right)+9\mathcal{F}_8(r)-3 \rho_0 \mathcal{F}_9(r)-4 \rho_0^3 \pi  r^2 \Big(403200 \rho_s^5 \nonumber\\&&\hspace{0.8cm} \times \pi ^2 \left(r^2-R^2\right)^3 \alpha ^4 r^{14}+4800 \rho_s^4 \pi ^2 R^2 \left(r^2-R^2\right) \alpha ^3 [21 (10 \beta -3) r^4+3 R^2 (41-140 \beta ) r^2+2 R^4 \nonumber\\&&\hspace{0.8cm} \times (105 \beta -26)] r^{12}+1200 \rho_s^3 \pi  R^4 \left(r^2-R^2\right) \alpha ^2 r^8 \mathcal{F}_{10}(r)+360 \rho_s^2 \pi  R^6 \left(r^2-R^2\right) \alpha r^6 \mathcal{F}_{11}(r)+2 \rho_s \nonumber\\&&\hspace{0.8cm} \times R^8 r^2 \mathcal{F}_{12}(r)-R^{10}  \times\mathcal{F}_{13}(r)\Big)+\rho_0^2 \mathcal{F}_4(r)\bigg],~~~~~~~~~~\label{eq2.2.23}
\end{eqnarray}
\end{small}
where the expressions for the coefficients used in the above expressions are given in the Appendix.

\subsection{Mimicking of the Density Constraint: $\rho(r)=\theta^0_0(r)$} \label{solB}

By mimicking the seed density ($\rho$) to the component ($\theta^0_0$) through Eqs. (\ref{eq19}) and (\ref{eq36}), we derive  
\begin{eqnarray}
  r \frac{d\mathcal{D}}{dr}  + \mathcal{D}   -\frac{8\pi r^2\left[r^2 (\rho_0-\rho_s)-\rho_0 R^2\right]}{R^2}=0.
\end{eqnarray}

After solving the above equation, we derive the exact solution of the decoupling function $\mathcal{D}(r)$ as
\begin{eqnarray}
    \mathcal{D}(r)=\frac{24 \pi  r^4 (\rho_0-\rho_s)-40 \pi  \rho_0 r^2 R^2}{15 R^2}.\label{eq2.12} 
\end{eqnarray}

Then, the deformed MGD solution for the density constraint approach can be expressed by the following spacetime geometries
\begin{eqnarray}
 && \hspace{-0.8cm}  e^{\Phi(r)}= \exp\Bigg[\frac{\sqrt{5}  \left(52 \pi  \alpha  \rho_0^2 R^2-45 \alpha  \rho_0+12 \pi  (\beta +1) \rho_0 R^2+9 \left(5 \alpha  \rho_s+8 \pi  \gamma  R^2\right)\right)}{12\sqrt{2 \pi } R \sqrt{\rho_0 \left(10 \pi  \rho_0 R^2-9\right)+9 \rho_s}}\nonumber\\&&\hspace{1.8cm}\times \tanh ^{-1}\left(\frac{\sqrt{\frac{2 \pi }{5}} \left(6 r^2 (\rho_s-\rho_0)+5 \rho_0 R^2\right)}{R \sqrt{\rho_0 \left(10 \pi  \rho_0 R^2-9\right)+9 \rho_s}}\right)+\frac{5 \alpha  r^2 (\rho_0-\rho_s)}{2 R^2}-\frac{1}{12} (15 \beta \nonumber\\&&\hspace{1.8cm} +5 \alpha  \rho_0+3)\log \left(24 \pi  r^4 (\rho_0-\rho_s)+5 R^2 \left(3-8 \pi  \rho_0 r^2\right)\right)+\mathcal{C}\Bigg]=e^{\Psi(r)}, \label{eq2.13}\\
  && \hspace{-0.8cm}    e^{-\Omega(r)}= \eta(r)+\sigma\, \mathcal{D} (r)=  \bigg[\frac{8}{15} \pi  \rho_0 r^2 \left(\frac{3 r^2}{R^2}-5\right)-\frac{8 \pi  \rho_s r^4}{5 R^2}+1\bigg]+\frac{\sigma}{{15 R^2}} \big[24 \pi  r^4 \nonumber\\&&\hspace{4.3cm} \times (\rho_0-\rho_s)-40 \pi  \rho_0 r^2 R^2\big] . \label{eq2.14} 
\end{eqnarray}

Now, the expressions for the components of new source $\theta_{ij}$ are written as: 
\begin{small}
\begin{eqnarray}
  && \hspace{-0.0cm}  \theta^0_0 (r)= \frac{\rho_0 \left(R^2-r^2\right)+\rho_s r^2}{R^2}, \label{eq2.15}\\
  && \hspace{-0.0cm}  \theta^1_1 (r)=\frac{\left(3 \rho_0 r^2-5 \rho_0 R^2-3 \rho_s r^2\right)}{R^4 \left(24 \pi  \rho_0 r^4-40 \pi  \rho_0 r^2 R^2-24 \pi  \rho_s r^4+15 R^2\right)}\Big[R^4 \big(8 \pi  \alpha  \rho_0^2 r^2+8 \pi  \beta  \rho_0 r^2\nonumber\\&&\hspace{0.7cm}+8 \pi  \gamma  r^2+1\big) +8 \pi  \alpha  r^6 (\rho_0-\rho_s)^2-8 \pi  r^4 R^2 (\rho_0-\rho_s) (\beta +2 \alpha  \rho_0)\Big],~~~~\label{eq2.16}\\
   && \hspace{-0.0cm}  \theta^2_2 (r)= \frac{1}{R^6 \left(-24 \rho_s \pi  r^4+8 \rho_0 \pi  \left(3 r^2-5 R^2\right) r^2+15 R^2\right)^2}\Big[-240 \pi ^2 \left(3 r^2-5 R^2\right) \left(r^3-r R^2\right)^4 \alpha ^2 \rho_0^5\nonumber\\&& \hspace{0.7cm}  +16 \pi ^2 r^4 \left(r^2-R^2\right) \alpha  \big(225 \rho_s \alpha  r^8-3 R^2 (265 \rho_s \alpha -30 \beta +18) r^6+3 R^4 (305 \rho_s \alpha -110 \beta +64) r^4 \nonumber\\&& \hspace{0.7cm}-5 R^6 (69 \rho_s \alpha -78 \beta +34) r^2-150 R^8 \beta \big) \rho_0^4 -4 \pi  r^2 \Big\{15 \left(24 r^6-71 R^2 r^4+62 R^4 r^2-15 R^6\right) \alpha  R^4\nonumber\\&& \hspace{0.7cm}+4 \pi  r^2 \big(450 \rho_s^2 \alpha ^2 r^{10}-18 \rho_s R^2 \alpha  (85 \rho_s \alpha -20 \beta +12) r^8+9 R^4  \big(190 \rho_s^2 \alpha ^2+2 \rho_s (41-70 \beta ) \alpha \nonumber\\&& \hspace{0.7cm}+10 \gamma  \alpha +5 \beta ^2-2 \beta +1\big) r^6-R^6 [630 \rho_s^2 \alpha ^2+4 \rho_s (181-360 \beta ) \alpha +330 \gamma  \alpha +165 \beta ^2-72 \beta +39] r^4\nonumber\\&& \hspace{0.7cm}+5 R^8 \left(39 \beta ^2-14 \beta +2 \rho_s \alpha  (17-54 \beta )+78 \alpha  \gamma +11\right) r^2-25 R^{10} \left(3 \beta ^2+6 \alpha  \gamma +1\right)\big)\Big\} \rho_0^3+4 \pi  r^2 \nonumber\\&& \hspace{0.7cm} \times \Big\{120 \rho_s^3 \pi   \left(15 r^4-34 R^2 r^2+19 R^4\right) \alpha ^2 r^8+8 \rho_s^2 \pi  R^2 \alpha  \big[54 (5 \beta -3) r^4-9 R^2 (70 \beta -41) r^2+R^4 \nonumber\\&& \hspace{0.7cm} \times (360 \beta -181)\big] r^6 +2 \rho_s R^4 \big(2 \pi \big\{27 \left(5 \beta ^2-2 \beta +10 \alpha  \gamma +1\right) r^4-6 R^2 \left(55 \beta ^2-24 \beta +110 \alpha  \gamma +13\right) \nonumber\\&& \hspace{0.7cm}  \times r^2+5 R^4 \left(39 \beta ^2-14 \beta +78 \alpha  \gamma +11\right)\big\} r^2+15 (36 r^4-71 R^2 r^2+31 R^4) \alpha \big) r^2+R^6 \big[72 \pi  (5 \beta +1) \nonumber\\&& \hspace{0.7cm} \times \gamma  r^6-6 \left(20 \pi  \gamma  R^2+5 \beta  \left(32 \pi  R^2 \gamma -9\right)+3\right) r^4+15 R^2 \big(\beta (40 \pi  R^2 \gamma  -37)+3\big) r^2+25 R^4 (9 \beta -1)\big]\Big\} \nonumber\\&& \hspace{0.7cm} \times \rho_0^2+\mathcal{G}_{1}(r)\Big]. \label{eq2.17}
\end{eqnarray}
where
\begin{eqnarray}
  &&\hspace{-0.0cm}  \mathcal{G}_{1}(r)=-3 \Big(80 \rho_s^4 \pi ^2 \left(15 r^2-17 R^2\right) \alpha ^2 r^{12}+32 \rho_s^3 \pi ^2 R^2 \alpha  \left(12 (5 \beta -3) r^2+R^2 (41-70 \beta )\right) r^{10}\nonumber\\&& \hspace{0.98cm} +4 \rho_s^2 \pi  R^4  \Big(36 \pi  \left(5 \beta ^2-2 \beta +10 \alpha  \gamma +1\right) r^4-4 \pi  R^2 \left(55 \beta ^2-24 \beta +110 \alpha  \gamma +13\right) r^2\nonumber\\&& \hspace{0.98cm} +5 \left(72 r^2-71 R^2\right) \alpha \Big) r^6+4 \rho_s \pi  R^6  \Big(48 \pi  (5 \beta +1) \gamma  r^4-4 \left(80 \pi  \beta  \gamma  R^2+10 \pi  \gamma  R^2-45 \beta +3\right) r^2\nonumber\\&& \hspace{0.98cm} +5 R^2 (3-37 \beta )\Big) r^4+5 R^8 \big[48 \pi ^2 \gamma ^2 r^6+16 \pi  \gamma   \left(3-5 \pi  R^2 \gamma \right) r^4 +\left(6-60 \pi  R^2 \gamma \right) r^2-5 R^2\big]\Big) \rho_0\nonumber\\&& \hspace{0.98cm} +18 \rho_s r^2 \Big(40 \rho_s^4 \pi ^2 \alpha ^2 r^{12}+16 \rho_s^3 \pi ^2 R^2 \alpha  (5 \beta -3) r^{10}+8 \rho_s^2 \pi  R^4 \big[\pi (5 \beta ^2-2 \beta +10 \alpha  \gamma +1) r^2\nonumber\\&& \hspace{0.98cm} +10 \alpha \big] r^6+4 \rho_s \pi  R^6 \left(4 \pi  \gamma  r^2+5 \beta  \left(4 \pi  \gamma  r^2+3\right)-1\right) r^4+5 R^8 \left(8 \pi ^2 \gamma ^2 r^4+8 \pi  \gamma  r^2+1\right)\big). \nonumber
\end{eqnarray}
\end{small}

\section{Boundary conditions for SS models under Gravitational decoupling}\label{sec4} 

To establish a realistic compact stellar model, characterized by a confined and bounded distribution of matter with a well-defined mass $M$ and radius $R$, it is essential to connect the inner geometry $\mathcal{M^{-}}$ at the surface $\Sigma=r=R$ with the outside space-time $\mathcal{M^{+}}$ that surrounds the structure. In the context of the GR system, the exterior manifolds is widely recognized as that of Schwarzschild vacuum space-time, specifically when considering uncharged, non-radiating, and static compact objects. However, it is necessary to examine the characteristics of the additional component of the energy-momentum tensor, namely the $\theta$-sector. Furthermore, this novel additional term has the potential to alter the material composition of the exterior space-time. The precise geometry that characterizes this exterior manifold can be provided as
\begin{equation}
ds^{2}=\left[1-\frac{2{\mathcal{M}}}{r}\right]dt^{2}- \frac{dr^2}{ 1-\frac{2{\mathcal{M}}}{r}+\sigma\, \mathcal{D^\ast}(r)} -r^{2}d\Omega^{2}. \label{eq3.1}
\end{equation}

The geometric deformation function for the Schwarzschild spacetime outside caused by the $\theta_{ij}$ source is called $\mathcal{D^\ast}(r)$. This metric (\ref{eq3.1}) denotes a distorted Schwarzschild space-time that is no longer a vacuum. Nevertheless, it is possible to make $\mathcal{D^\ast}(r)$ null in order to preserve the typical outer vacuum space-time, without compromising its generality and benefiting from simplicity.  To effectively integrate the internal configuration with the external one, the ID junction conditions need the use of both the first and second basic forms. The first basic form determines the continuity of the metric potentials at the $\Sigma$ boundary. The first basic form is as follows:
\begin{equation}\label{eq3.2}
e^{\Phi^{-}(r)}|_{r=R}=e^{\Phi^{+}(r)}|_{r=R},
\end{equation}
and
\begin{equation}\label{eq3.3}
e^{\Psi^{-}(r)}|_{r=R}=e^{\Psi^{+}(r)}|_{r=R},
\end{equation}
where the symbols $``-"$ and $``+"$ represent the inner and outer geometries, respectively. 

The second basic form refers to the continuity of the extrinsic curvature $K_{\mu\nu}$ caused by the components $\mathcal{M}^{-}$ and $\mathcal{M}^{+}$ on the surface. The continuity of the $K_{rr}$ component throughout the field $\Sigma$ results in
\begin{equation}\label{eq3.4}
\left[P_r^{(\text{eff})}(r)\right]_{\Sigma}=\left[P^{(\text{eff})}_r(r)-\sigma\, \theta^{1}_{1}(r)\right]_{\Sigma}=0.
\end{equation} 
which yields
  \begin{equation}
{P_r}(R)+\sigma\,(\theta^1_1)^{-}(R)=\sigma\,(\theta^1_1)^{+}(R).~~~\label{eq3.5}
  \end{equation}
  
Referring to Eqs. (\ref{eq37}) and (\ref{eq3.5}), we may now deduce the following result:
\begin{eqnarray}
{P}_r(R)+\sigma\,\bigg[\mathcal{D}(R)\left(\frac{\Psi^{\prime}(R)}{R}+\frac{1}{R^{2}}\right)\bigg] =\sigma\,(\theta^1_1)^{+}(R).~~\label{eq3.6}
\end{eqnarray}

The substitution of the outside space-time in Eq. (\ref{eq37}) with Eq. (\ref{eq3.6}) yields
\begin{eqnarray}
 P_r(R)+\sigma\,\bigg[\mathcal{D}(R)\left(\frac{\Psi^{\prime}(R)}{R}+\frac{1}{R^{2}}\right)\bigg] =\sigma\,\mathcal{D}^{\ast}(R)\Bigg[\frac{2\mathcal{M}}{R^3\,\Big(1-\frac{2\mathcal{M}}{R}\Big)}+\frac{1}{R^2}\Bigg].~~~\label{eq3.7}
 \end{eqnarray} 
 
The Eqs. (\ref{eq3.2}), (\ref{eq3.3}) and (\ref{eq3.7}) are necessary conditions for the establishment of a spherically symmetric external ``vacuum" specified by the deformed Schwarzschild-de Sitter metric in Eq. (\ref{eq3.1}) with respect to the interior MGD metric~(\ref{eq4}). The external structure of the system could be comprised of fields that are defined in the source $\theta_{ij}$. A fundamental second form is derived from the matching condition (\ref{eq3.7}): if the outer geometry is defined by the exact Schwarzschild metric, then it is necessary to have $\mathcal{D}^{\ast}(r)=0$ in Eq.~(\ref{eq3.1}), which therefore results in the condition.
\begin{eqnarray}
 P^{\text{eff}}_r(R)=P_r(R)+\sigma\,\bigg[\mathcal{D}(R)\left(\frac{\Psi^{\prime}(R)}{R}+\frac{1}{R^{2}}\right)\bigg] =0.~~ \label{eq3.8} 
 \end{eqnarray}  
 
Now we shall use conditions (\ref{eq3.2}), (\ref{eq3.3}) and (\ref{eq3.8}) to derive the arbitrary constant involved in the solution.  Through these conditions, the expressions for the constants corresponding to solutions (\ref{solA}) and (\ref{solB}) are given in the following subsections.

\subsection{The expressions of the constant for the solution~\ref{solA}:}
\begin{small}
  \begin{eqnarray}
&&\hspace{-0.0cm}    \gamma=-\rho_s (\beta +\alpha  \rho_s),\\
&&\hspace{-0.0cm}    \mathcal{C}= \frac{1}{24} \Bigg[\frac{\sqrt{\frac{10}{\pi }} \tanh ^{-1}\left(\frac{\sqrt{\frac{2 \pi }{5}} R (\rho_0-6 \rho_s)}{\sqrt{10 \pi  \rho_0^2 R^2-9 \rho_0+9 \rho_s}}\right) \left(52 \pi  \alpha  \rho_0^2 R^2-45 \alpha  \rho_0+12 \pi  (\beta +1) \rho_0 R^2+9 \left(5 \alpha  \rho_s+8 \pi  \gamma  R^2\right)\right)}{R \sqrt{10 \pi  \rho_0^2 R^2-9 \rho_0+9 \rho_s}}\nonumber\\&&\hspace{0.8cm}+24 \log \left(\frac{\left(16 \pi  \rho_0 R^2+24 \pi  \rho_s R^2-15\right) \left(8 \pi  \alpha  \rho_s^2 R^2 (\sigma -1)+8 \pi  \beta  \rho_s R^2 (\sigma -1)+8 \pi  \gamma  R^2 (\sigma -1)-1\right)}{15 \left(8 \pi  \alpha  \rho_s^2 R^2+8 \pi  \beta  \rho_s R^2+8 \pi  \gamma  R^2+1\right)}\right)\nonumber\\&&\hspace{0.8cm}-60 \alpha  (\rho_0-\rho_s)+2 (15 \beta +5 \alpha  \rho_0+3) \log \left(15 R^2-8 \pi  R^4 (2 \rho_0+3 \rho_s)\right)\Bigg].
\end{eqnarray}
\end{small}

\subsection{The expressions of the constant for the solution~\ref{solB}:}
 \begin{small}
\begin{eqnarray}
&&\hspace{-0.0cm}    \gamma=-\frac{1}{16 \pi  \rho_0 R^2 (\sigma +1)+24 \pi  \rho_s R^2 (\sigma +1)-15} \Big[\rho_s^2 \left(\alpha  \left(16 \pi  \rho_0 R^2 (\sigma +1)-15\right)+24 \pi  \beta  R^2 (\sigma +1)\right)+\rho_s \nonumber\\&&\hspace{0.8cm} \times \left(\beta  \left(16 \pi  \rho_0 R^2 (\sigma +1)-15\right)+3 \sigma \right)+2 \rho_0 \sigma +24 \pi  \alpha  \rho_s^3 R^2 (\sigma +1)\Big],~~~~~\\
&&\hspace{-0.0cm}    \mathcal{C}= \frac{1}{24} \Bigg[\frac{\sqrt{\frac{10}{\pi }} \tanh ^{-1}\left(\frac{\sqrt{\frac{2 \pi }{5}} R (\rho_0-6 \rho_s)}{\sqrt{10 \pi  \rho_0^2 R^2-9 \rho_0+9 \rho_s}}\right) \left(52 \pi  \alpha  \rho_0^2 R^2-45 \alpha  \rho_0+12 \pi  (\beta +1) \rho_0 R^2+9 \left(5 \alpha  \rho_s+8 \pi  \gamma  R^2\right)\right)}{R \sqrt{10 \pi  \rho_0^2 R^2-9 \rho_0+9 \rho_s}}\nonumber\\&&\hspace{0.8cm} -60 \alpha  (\rho_0-\rho_s)+24 \log \left(-\frac{1}{15} 16 \pi  (\beta +1) \rho_0 R^2-\frac{8}{5} \pi  (\beta +1) \rho_s R^2+1\right)+2 (15 \beta +5 \alpha  \rho_0+3) \nonumber\\&&\hspace{0.8cm} \times \log \left(15 R^2-8 \pi  R^4 (2 \rho_0+3 \rho_s)\right)\Bigg].
\end{eqnarray}
 \end{small}

\section{Physical analysis of minimally deformed solutions in strange star models}\label{sec5} 

We investigate important thermodynamic variables -- energy density, radial and tangential stresses, and the anisotropic parameter -- in great depth here.
With an emphasis on two distinct solutions (see Sect. \ref{sec3}) -- the pressure constraint modeled by $P_r(r) = \theta^1_1(r)$~[\ref{solA}] and the density constraint modeled by $\rho(r) = \theta^0_0(r)$~[\ref{solB}] this work aims to clarify the relevance of our deformed SS models in describing astrophysical events.

\subsection{Physical behavior of energy density and pressure profiles}\label{sec5.1} 

\begin{figure*}
\centering
\includegraphics[height=6cm,width=7.5cm]{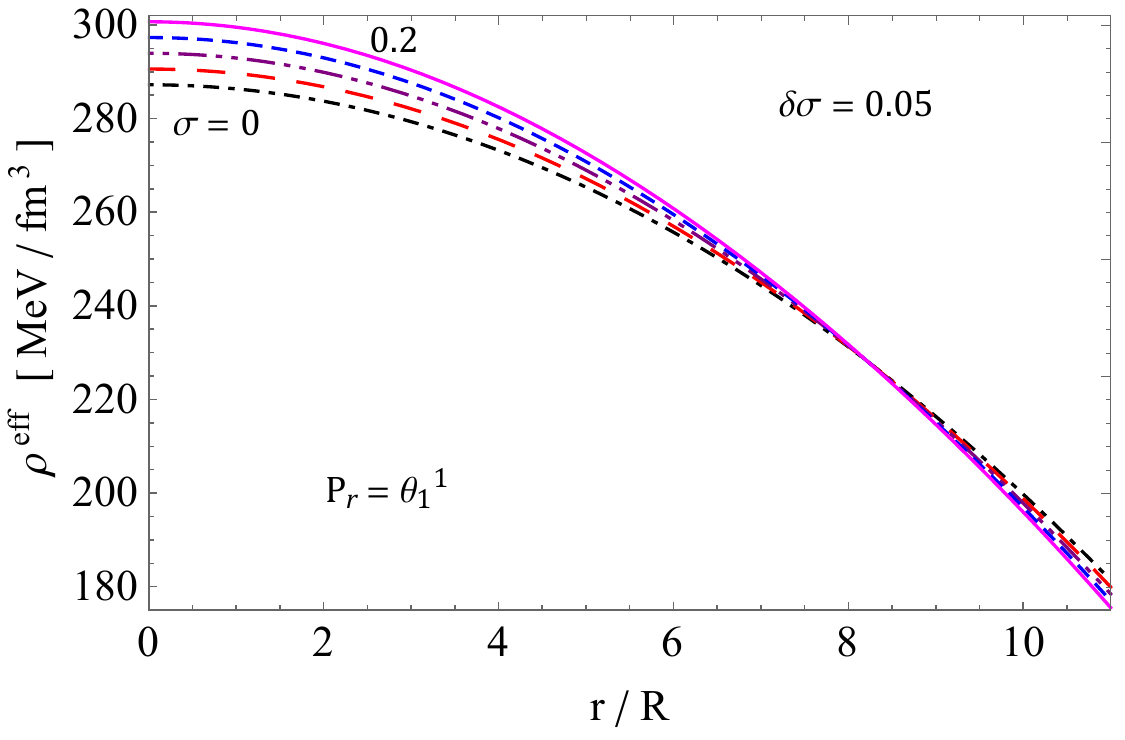}~~~~~
\includegraphics[height=6cm,width=7.5cm]{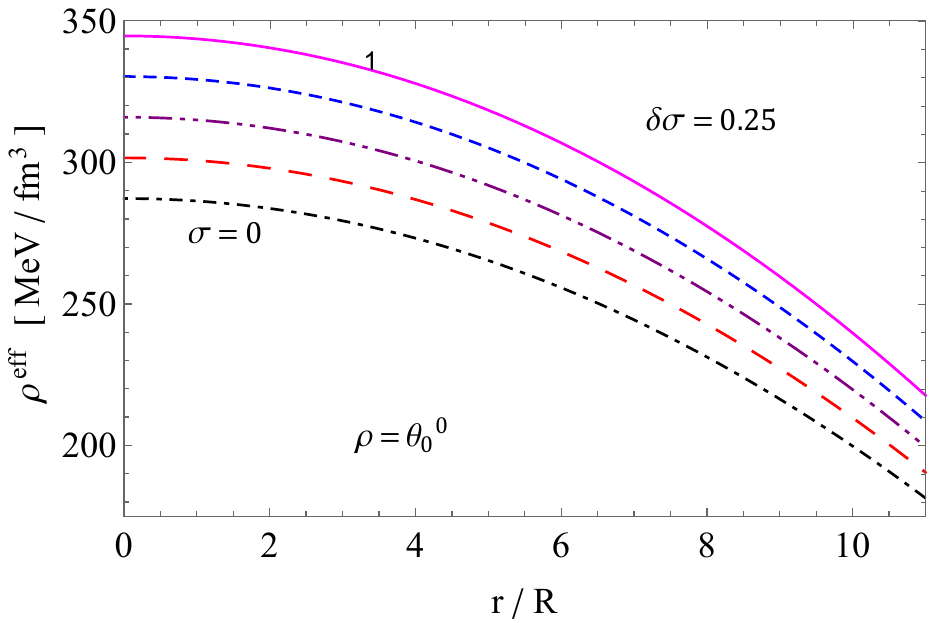}
\caption{Decoupling effect on energy density ($\rho^{\text{eff}}$) against radial coordinate $r$ for the solution~\ref{solA} (left panel: $\alpha =100,\beta =0.15,\rho_{_0}=0.00038;\rho_{_s}=0.00024$) and solution~\ref{solB} (right panel: $\alpha =100,\beta =0.15,\rho_{_0}=0.00038;\rho_{_s}=0.00024$).}
\label{f1}
\end{figure*}

The effective energy density, which we refer to as $\rho^{\text{eff}}$, shows a fascinating pattern in the context of deformed SS models. At the center of these stars, the energy density is at its highest point. As we move outward toward the surface, this density gradually decreases, reaching its lowest value right at the surface. We can see this trend illustrated in Fig. \ref{f1}. This figure examines how the decoupling effect, indicated by $\sigma$, influences the effective energy density ($\rho^{\text{eff}}$) at different radial coordinates $r$. In the left panel, which corresponds to solution -- $P_r(r) = \theta^1_1(r)$~[\ref{solA}], the values of $\sigma$ range from 0 to 0.2. Meanwhile, the right panel, related to the solution -- $\rho(r) = \theta^0_0(r)$~[\ref{solB}], shows a range of 0 to 1. Both panels confirm that the effective energy density remains stable and well-defined at all points within the star for both solutions. This regularity highlights how robust the effective energy density is across various radial coordinates, emphasizing the important role that the decoupling effect plays in shaping the energy distribution within these intriguing stellar models.

In the first solution -- $P_r(r) = \theta^1_1(r)$~[\ref{solA}], we notice that the core density is relatively low. On the other hand, the second solution---$\rho(r) = \theta^0_0(r)$~[\ref{solB}] shows a slight increase in core density. This difference is mainly due to the decoupling effect, denoted $\sigma$. As $\sigma$ increases, the density in the central regions of the star also increases a bit, leading to the expansion of matter into concentric shells. What is interesting is that the impact of $\sigma$ becomes even more noticeable when we look at the transition from the core to the outer layers of the star. In both solutions, variations in $\sigma$ play a significant role in shaping the overall density of the star. It is worth noting that a higher decoupling effect $\sigma$ increases the gravitational pull on matter at higher densities. This causes the matter density to shift towards higher equilibrium values throughout the star's interior. This relationship highlights how the decoupling effect influences the distribution of density in these fascinating stellar models.

Examining the effective radial ($P_r^{\text{eff}}$) and tangential ($P_t^{\text{eff}}$) pressures reveals some fascinating differences between the two solutions. In the first solution -- $P_r(r) = \theta^1_1(r)$~[\ref{solA}], we see that the effective pressures in the central region are noticeably lower than in the second solution -- $\rho(r) = \theta^0_0(r)$~[\ref{solB}]. Although the magnitudes differ, both solutions exhibit similar trends as we move outward from the center. A significant aspect of both solutions is that the effective radial and tangential pressures remain continuous throughout the star. As we approach the surface, both pressures gradually decrease, with the radial pressure eventually vanishing altogether at the stellar surface. This shift is an important feature of the star's overall structure. The role of the decoupling effect, $\sigma$, is particularly interesting. When $\sigma$ increases from 0 to 0.2 in the first solution -- $P_r(r) = \theta^1_1(r)$~[\ref{solA}], we notice a clear reduction in the effective radial and tangential pressures in the central region, although these pressures were already lower to begin with. On the flip side, in the second solution  $\rho(r) = \theta^0_0(r)$~[\ref{solB}], raising $\sigma$ from 0 to 1 leads to an increase in both types of effective pressure in the core. This pattern highlights how the decoupling effect $\sigma$ influences the fluid particles within the star. Higher values of $\sigma$ create a confining and compacting effect, especially in the core regions, illustrating the complex interactions at work in these stellar models.

\subsection{Physical behavior of anisotropic profiles}\label{sec5.2} 

To truly understand the stability of a star, we need to look closely at the anisotropy parameter, $\Delta^{\text{eff}}$. As shown in Fig. \ref{f3}, this parameter starts at zero at the center of the star and gradually increases as we move toward the outer boundary. In the first solution \ref{solA}, represented by $P_r(r) = \theta^1_1(r)$, we notice that $\Delta^{\text{eff}}$ actually reaches its highest point just before hitting the boundary. This upward trend tells us that the effective radial stresses are stronger than the transverse stresses, resulting in an outward repulsive force. This growing anisotropy is not just a technical detail; it plays a vital role in keeping the star's outer layers stable. By pushing back against the gravitational forces that want to pull everything inward, this increasing anisotropy and its associated repulsive forces help to maintain the overall stability of the star. The interplay between the rising anisotropy, the repulsive forces it creates, and the balance with gravitational attraction is crucial for the star's structural integrity. This delicate balance not only supports the surface layers but also strengthens the star's ability to resist collapse, highlighting just how important anisotropic pressure is for stellar stability. 

It is important to highlight a significant difference between the two solutions we are examining. In the first solution \ref{solA} ($P_r(r) = \theta^1_1(r)$), the behavior of the anisotropy parameter is quite different from what we see in the second solution \ref{solB} ($\rho(r) = \theta^0_0(r)$). As the parameter $\sigma$ increases in the second solution, we observe a dramatic surge in anisotropy, about ten times higher than in the first solution. This stark contrast really shows how much $\sigma$ influences the anisotropic behavior within the star. This boost in anisotropy is not just a simple change; it is the result of several intricate physical processes happening inside the star. For instance, the movement of neutrinos and electrons significantly alters the pressure distribution, leading to varying levels of anisotropy throughout the stellar structure. Additionally, phase transitions can change the EOS, which affects the balance between radial and tangential stresses. In addition, dissipative effects such as viscosity and heat conduction also play a key role in creating anisotropic stresses. Together, these factors create a dynamic environment within the star that greatly enhances its anisotropy. Understanding these complex interactions is crucial because they not only increase the anisotropic behavior of the star but also have important implications for its stability and evolution over time.

\begin{figure*}
\centering
\includegraphics[height=6cm,width=7.5cm]{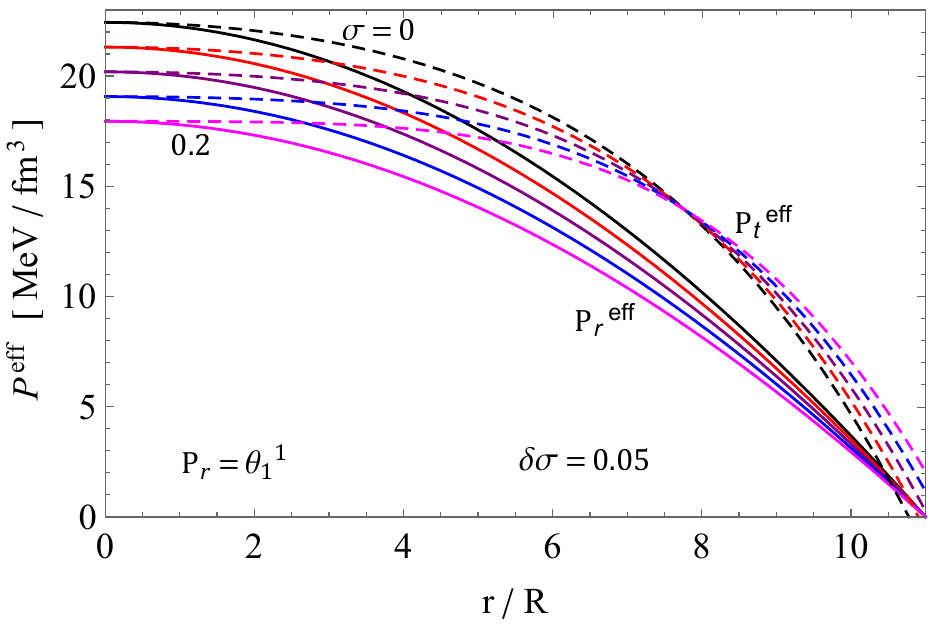}~~~~~
\includegraphics[height=6cm,width=7.5cm]{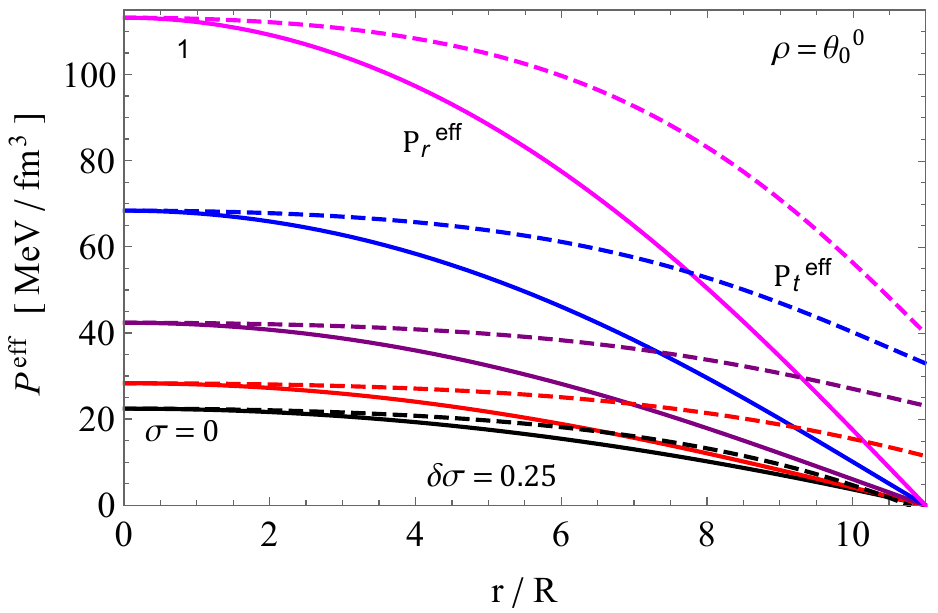}
\caption{Decoupling effect on radial and tangential pressures ($p^{\text{eff}}_r-\text{solid~lines~and}~ p^{\text{eff}}_t-\text{dashed~lines}$)  against radial coordinate $r$ for the solution~\ref{solA} (left panel) and solution~\ref{solB} (right panel).}
\label{f2}
\end{figure*}

\begin{figure*}
\centering
\includegraphics[height=6cm,width=7.5cm]{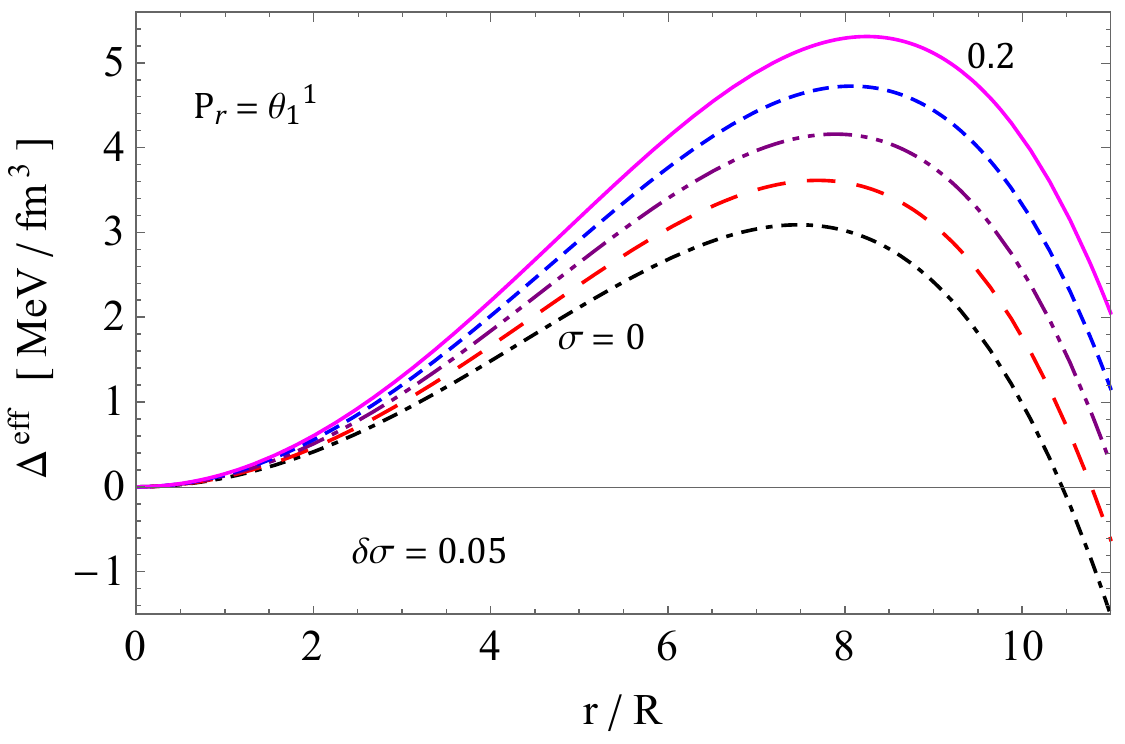}~~~~~
\includegraphics[height=6cm,width=7.5cm]{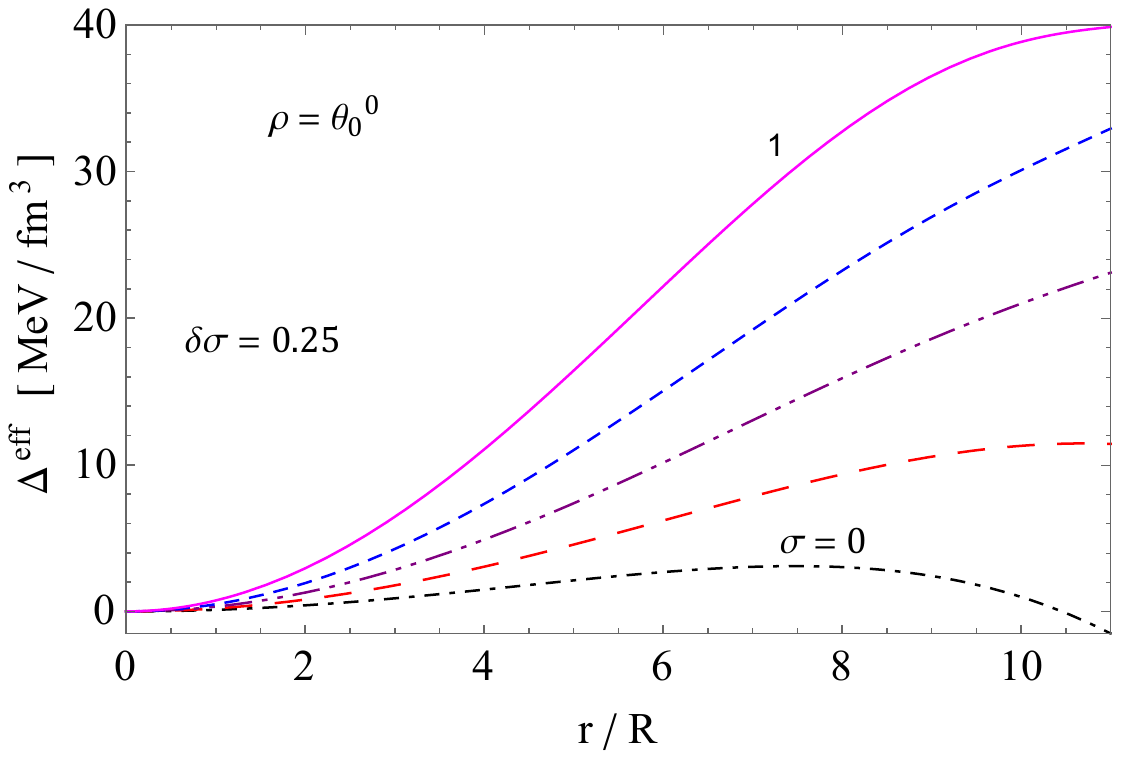}
\caption{Decoupling effect on anisotropy ($\Delta^{\text{eff}}$  against radial coordinate $r$ for the solution~\ref{solA} (left panel) and solution~\ref{solB} (right panel).}
\label{f3}
\end{figure*}

\section{Stability analysis}\label{sec6}

In this section, we are going to dive into the stability of stellar structures using two well-known methods. First, we have the adiabatic index, which helps us understand how a star reacts to changes in the pressure and density -- basically, how stable it is when things get a little shaken up. After that, we shall take a look at the Harrison-Zel'dovich-Novikov criteria, another popular approach for assessing the stability of compact stars. This method considers important factors like the EOS and how mass is distributed within the star. By using these two approaches, we aim to get a clearer picture of how stable the stellar objects we are studying really are. Each method offers unique insights, and together, they'll help us form a well-rounded understanding of stability in these fascinating systems.

\subsection{Stability analysis via adiabatic index} \label{sec6.1}

Here, we are going to take a closer look at the stability of anisotropic stellar configurations. To do this, we need to analyze something called the adiabatic index, which we denote as $\Gamma$. This index helps us understand how the pressure and density within a star relate to each other, and it's defined by the following equation:
\begin{equation}
    \Gamma = \frac{\rho + P_r^{\text{eff}} }{P_r^{\text{eff}}} \frac{dP_r^{\text{eff}}}{d\rho^{\text{eff}}}.
\end{equation}

For isotropic fluids in a Newtonian framework, there is a simple stability condition: the adiabatic index $\Gamma$ must be greater than $\frac{4}{3}$. This criterion has been discussed in the following works \citep{heintzmann1975neutron,bondi1992anisotropic}, especially in relation to neutron stars and anisotropic fluids. However, when we switch our focus to anisotropic stellar models, the stability requirements become a bit more complex than what we see with isotropic fluids, according to the classic results from Chandrasekhar \citep{chan1992dynamical,chan1993dynamical}. The stability condition for anisotropic stars is given by:
\begin{equation}
\Gamma > \frac{4}{3}\left(1 + \frac{\Delta^{\text{eff}}}{r|(P_{r}^{\text{eff}})|^{\prime}} + \frac{1}{4} \frac{8\pi \rho^{\text{eff}} P_{r}^{\text{eff}}r}{|(P_{r}^{\text{eff}})|^{\prime}} \right).\label{eq62}
\end{equation}
In this equation, the prime symbol indicates that we are differentiating with respect to the radial coordinate, $r$. The second term accounts for how anisotropy (where $P_r$ is not equal to $P_t$) changes the stability condition, while the last term adds in some relativistic corrections. Now, let us talk about the adiabatic index itself. The value of $\Gamma$ must meet certain criteria to ensure that an isotropic fluid sphere remains stable. This critical threshold is known as the critical adiabatic index, $\Gamma_{cr}$, and it's defined by the inequality:
\begin{equation}
\langle \Gamma \rangle > \Gamma_{cr},
\end{equation}
where $\langle \Gamma \rangle$ is the average adiabatic index \citep{Moustakidis:2016ndw}. 

According to previous studies \citep{Moustakidis:2016ndw,Koliogiannis:2018hoh}, we can express the critical value as:
\begin{equation}
\Gamma_{cr} = \frac{4}{3} + \frac{19}{42} \mathcal{U}.
\end{equation}

Here, $\mathcal{U} = \frac{2M}{R}$ represents the compactness parameter in the context of GR. In a purely Newtonian framework, this critical value stays constant at $\Gamma_{cr} = \frac{4}{3}$. However, when we consider the effects of GR, this critical value actually increases beyond $\frac{4}{3}$, reflecting the additional complexities that arise in relativistic scenarios.

\begin{figure*}
\centering
\includegraphics[height=6cm,width=7.5cm]{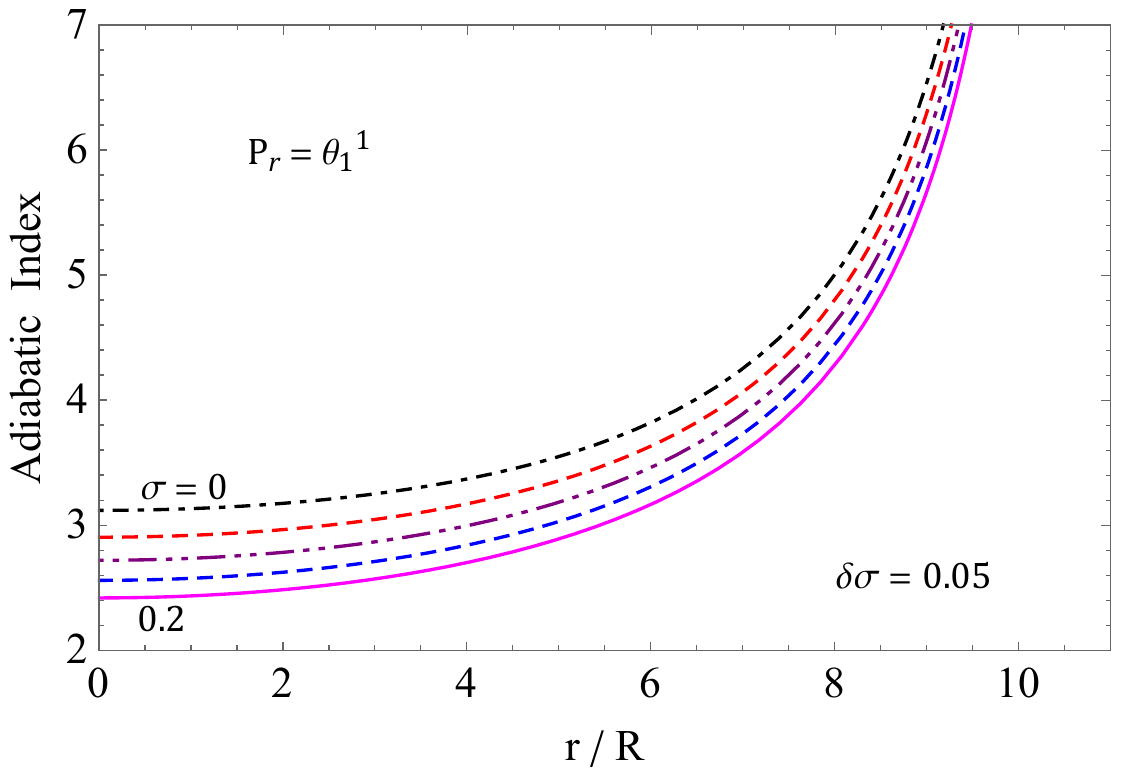}~~~~~
\includegraphics[height=6cm,width=7.5cm]{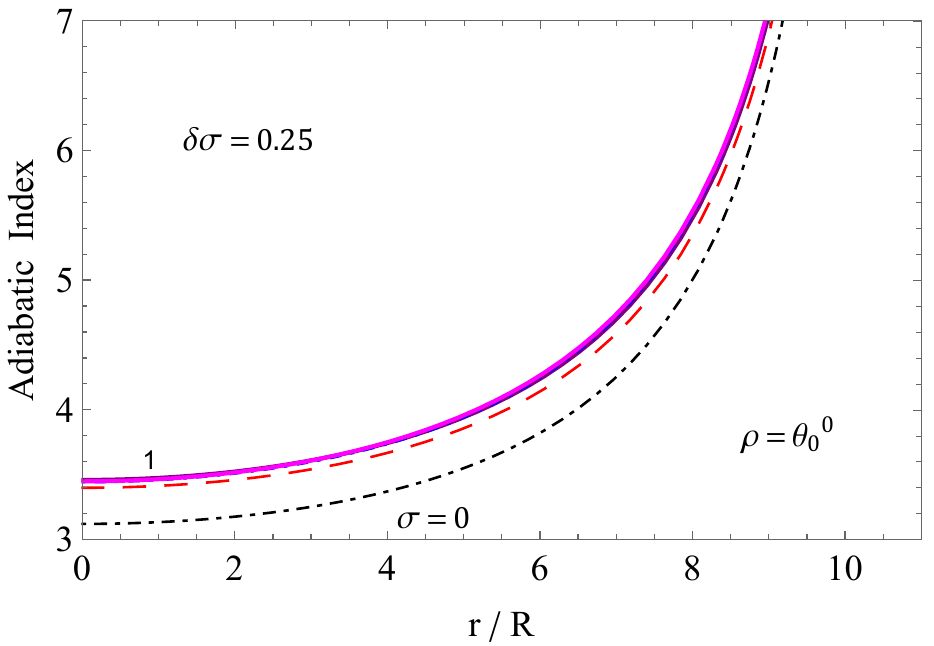}
\caption{Decoupling effect on adiabatic index ($\Gamma$) against the radial coordinate $r$ for the solution~\ref{solA} (left panel) and solution~\ref{solB} (right panel).}
\label{f4}
\end{figure*}

To better understand how the adiabatic index $\Gamma$ behaves, let us take a look at Fig. \ref{f4}. This figure uses the same sets of parameters that we discussed in Figs. \ref{f1} to \ref{f3}, making it easier to see how everything connects. What we find in Fig. \ref{f4} is quite interesting: there is a clear link between the adiabatic index and the stability of the star's structure. Specifically, we see that the two solutions show a strong correlation with stable configurations. The first solution relates to the pressure inside the star, which we can describe with the equation $P_r(r) = \theta^1_1(r)$~[\ref{solA}]. This indicates that the way pressure changes with radius is crucial for maintaining stability. The second solution focuses on the density of the star, represented by $\epsilon(r) = \theta^0_0(r)$~[\ref{solA}]. This shows us that the density distribution also plays an important role in ensuring that the star remains stable. In short, these results highlight how both pressure and density are key players in the stability of a star. By analyzing the adiabatic index in this way, we gain valuable insights into how these factors work together to form the dynamics of stellar structures.

\subsection{Stability analysis via cracking criterion}\label{sec6.2}

A key aspect of looking into how sound speeds change within a star is figuring out its equilibrium state. This state can be stable or unstable, according to the cracking concept presented by \cite{Herrera1992}. This framework helps us pinpoint stable regions in the star and those that might be at risk of instability, depending on the following conditions:
\begin{eqnarray}
     \label{169}
 -1\leq v^{2}_{t}-v^{2}_{r}\leq 1 \hspace{0cm}= 
 \left\{ \begin{array}{ll}
		   -1\leq v^{2}_{t}-v^{2}_{r}\leq 0~~~& \mathrm{Potentially\ stable\ }  \\
		 0< v^{2}_{t}-v^{2}_{r}\leq 1~~~& \mathrm{Potentially\ unstable}
	       \end{array}
	     \right\},~~~
   \end{eqnarray}
   
\begin{figure*}
\centering
\includegraphics[height=6cm,width=7.5cm]{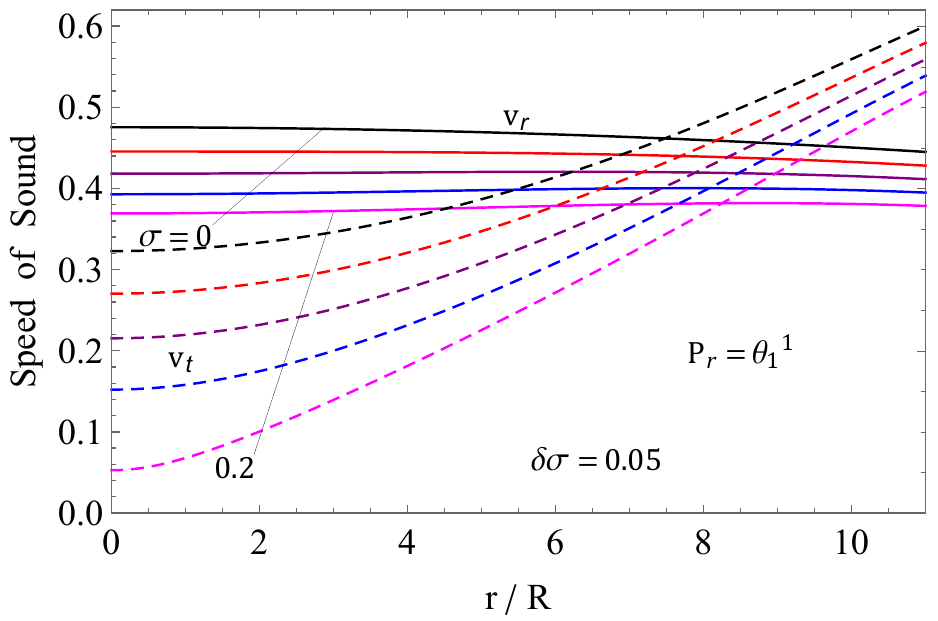}~~~~~
\includegraphics[height=6cm,width=7.5cm]{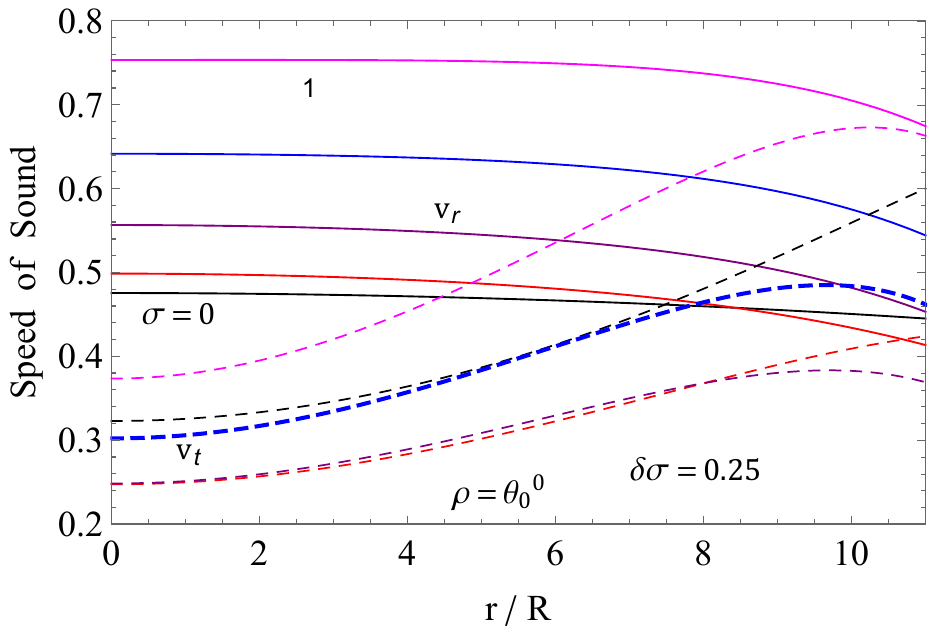}
\caption{Decoupling effect on the radial and tangential speed of sounds ($v^{2}_r-\text{solid~lines~and}~ v^{2}_t-\text{dashed~lines}$) against radial coordinate $r$ for the solution~\ref{solA} (left panel) and solution~\ref{solB} (right panel).}
\label{f5}
\end{figure*}

First, we took a deep dive into whether our current model upholds the principle of causality. To make this clear, we created a graph that shows how the squares of the sound speeds vary in both the radial and tangential directions as a function of the radial coordinate $r$. We can see this feature in Fig. \ref{f5} in an illustrative manner under the decoupling effect on the radial and tangential speed of sounds against the radial coordinate for the solutions 3.1 and 3.2, respectively. The graph clearly exhibits that both the sound speeds ($v^{2}_{r}$ and $v^{2}_{t}$) stay below the speed of sound, essentially they remain below unity, across all increasing values of $\sigma$ in both solution~\ref{solA} (left panel) and solution~\ref{solB} (right panel). This strongly suggests that the influences of MGD within GR do not compromise causality in any way. Additionally, by integrating the modified density model suggested by Mak \& Harko~\cite{Harko:2002prx} with MGD and a generalized quadratic EOS, we discovered that the sound speeds meet Herrera's inequality within the stellar system. This requirement is crucial to ensure stable equilibrium, highlighting the importance of MGD in preserving the robustness of our model.

\subsection{Stability analysis via Harrison-Zel'dovich-Novikov criteria}\label{sec6.3}

To better understand the connection between the stellar mass MM and the central energy density $\rho^{\text{eff}}_0$, we turn to the static stability criterion \citep{ZHN1, ZHN2}. Our exploration will focus on two different approaches: one that looks at the pressure constraint, given by $P_r(r) = \theta^1_1(r)$~[\ref{solA}], and another that examines the density constraint, defined by $\rho(r) = \theta^0_0(r)$~[\ref{solB}] 
\begin{eqnarray}
\frac{dM}{d\rho^{\text{eff}}_0} < 0 &\hspace{0.5cm}& \rightarrow \mbox{the configuration is considered unstable},\ \\
\frac{dM}{d \rho^{\text{eff}}_0} > 0 &\hspace{0.5cm}& \rightarrow \mbox{the configuration is deemed stable}.
\end{eqnarray}

\begin{figure}[!htp]
    \centering
\includegraphics[height=6.2cm,width=8cm]{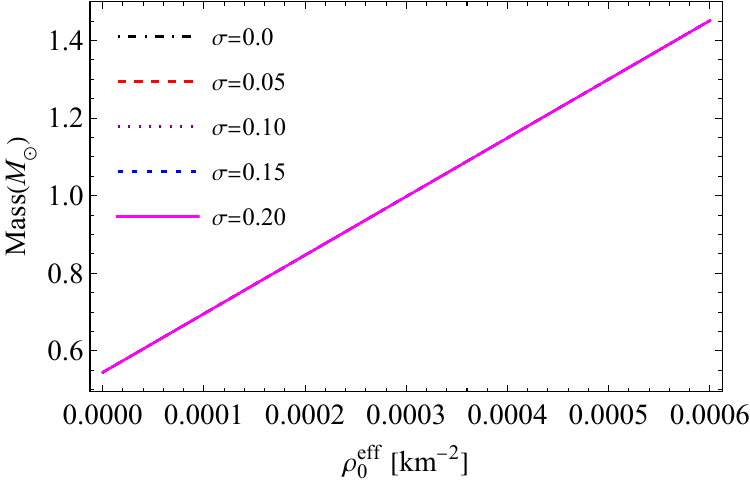}~~~\includegraphics[height=6.2cm,width=8cm]{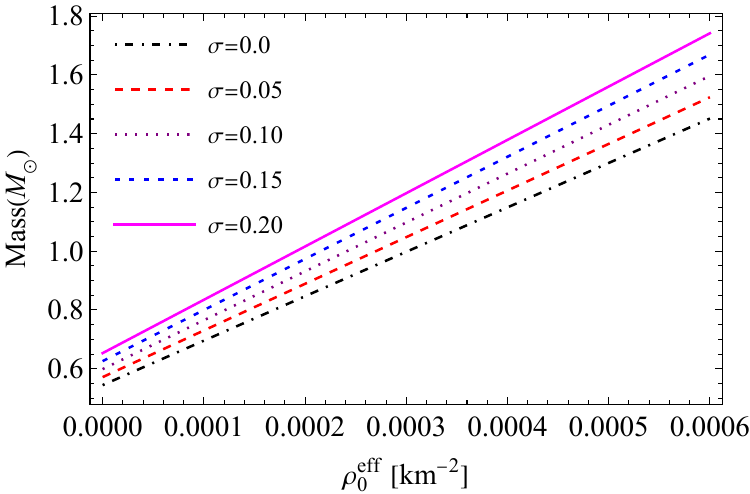}
    \caption{Stability analysis via $M-\rho^{\text{eff}}_0$ within the stellar object corresponding to different values of decoupling parameter $\sigma$ with fixed parameter values as used in Fig.~\ref{f4} for the solution~\ref{solA} (left panel) and solution~\ref{solB} (right panel).}
    \label{fig5a}
\end{figure}

These rules hold true for all stellar configurations, where $\rho^{\text{eff}}_0$ refers to the effective central density of the seed system. From the analysis in Fig. \ref{fig5a}, it is clear that $M(\rho^{\text{eff}}_0)$ is both positive and increasing, indicating that $\frac{dM}{d\rho^{\text{eff}}_0}$ is positive. This points to the stability of the stellar structures that are formed. 

Interestingly, within the same range of fluctuations in $\rho^{\text{eff}}_0$, stellar objects with different values of the decoupling parameter $\sigma$ show a slight increase in mass, with a matched trend for all the chosen values of $\sigma$, especially for the solution based on the pressure constraint $P_r(r) = \theta^1_1(r)$~[\ref{solA}]. On the other hand, the solution that uses the density constraint $\rho(r) = \theta^0_0(r)$~[\ref{solB}] exhibits a significant change when $\sigma$ varies from 0 to 0.2, suggesting that stability improves with small density variations.

\begin{figure*}
\centering
\includegraphics[height=6.2cm,width=8cm]{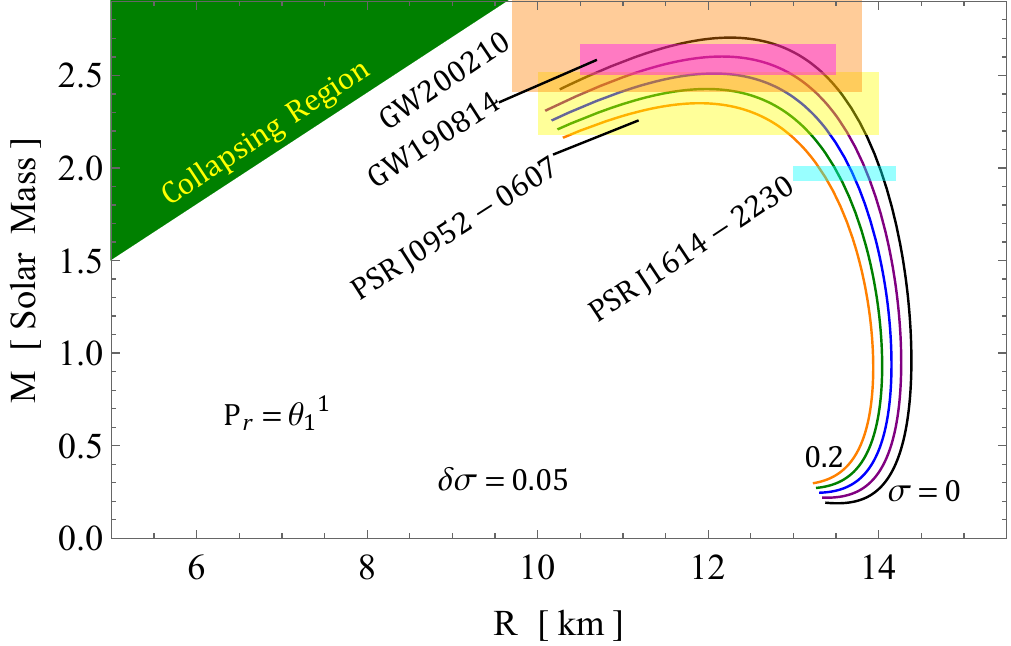}~~~~~
\includegraphics[height=6.2cm,width=8cm]{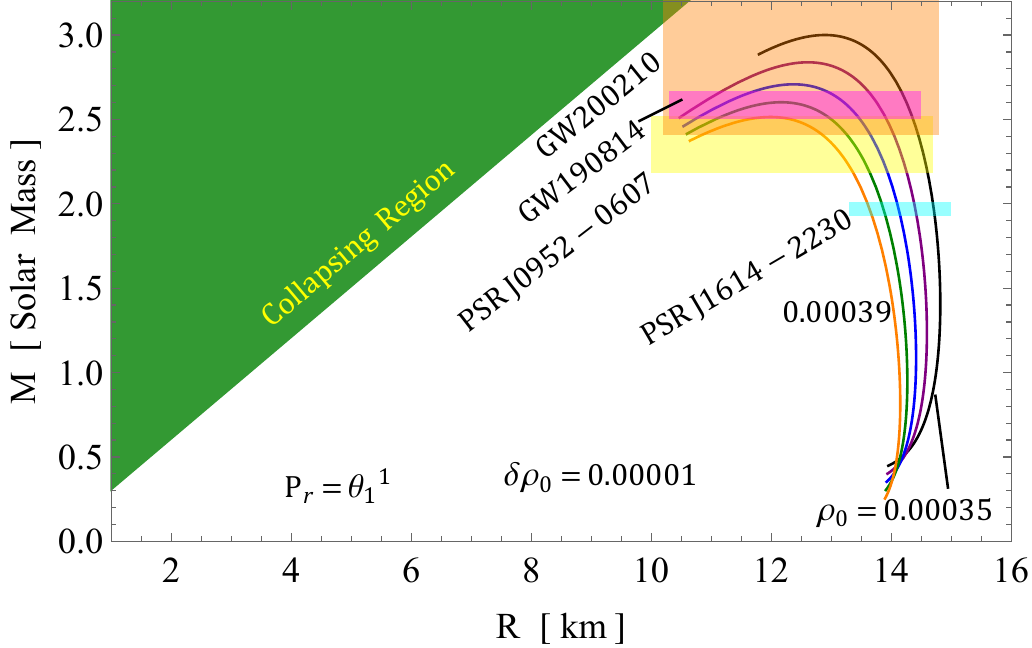}
\caption{$M-R$ curves for different values of $\sigma$ and $\rho_{_0}$ for $P_r$ mimicking $\theta_1^1$ for $\alpha = 100,~ \beta = 0.15, ~ \rho_{_s} =0.00024/km^2$ and $\alpha =100,~\beta =0.15,~\rho_{_s}=0.00024/km^2,~\sigma =0.05$. }
\label{f6}
\end{figure*}

\begin{figure*}
\centering
\includegraphics[height=6.2cm,width=8cm]{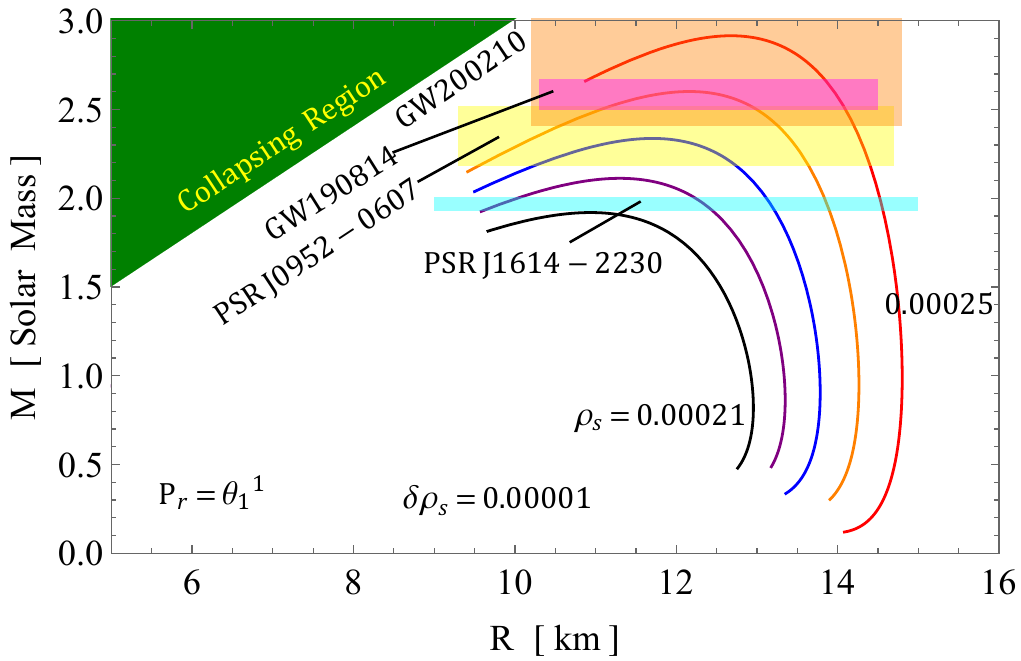}~~~~~
\includegraphics[height=6.2cm,width=8cm]{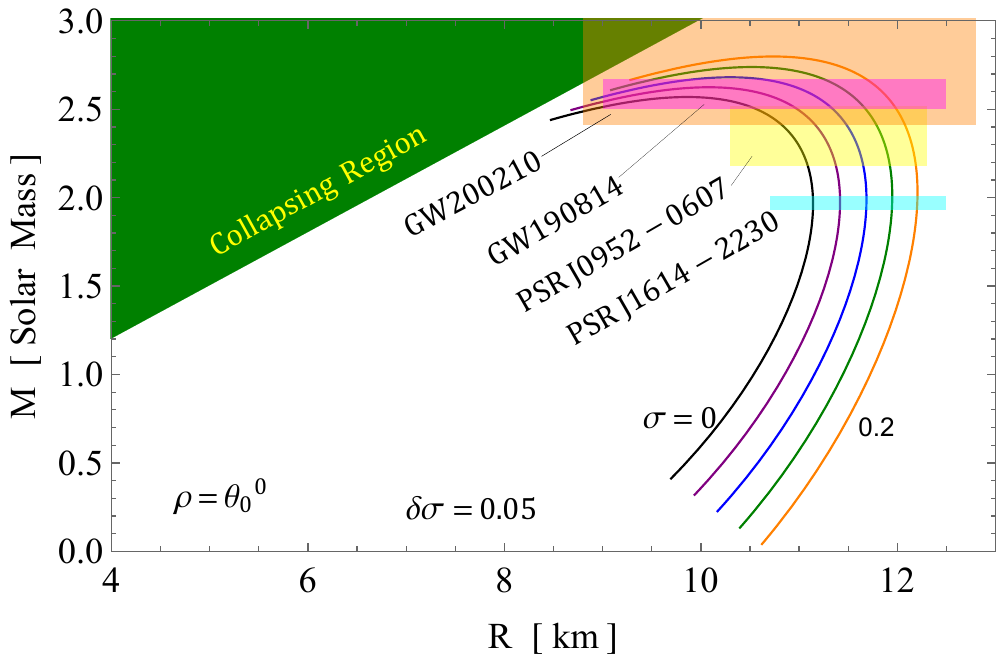}
\caption{$M-R$ curves for different values of $\sigma$ and $\rho_{_s}$ for $P_r$ mimicking $\theta_1^1$ and $\rho=\theta_0^0$ for $\alpha = 100,~ \beta = 0.15,~ \rho_{_0} =0.00038/km^2,~\sigma =0.05$ and $\alpha =100,~\beta =0.15,~\rho_{_0} = 0.00038/km^2,~\rho_{_s} = 0.00024/km^2$.}
\label{f7}
\end{figure*}

\begin{figure*}
\centering
\includegraphics[height=6.2cm,width=8cm]{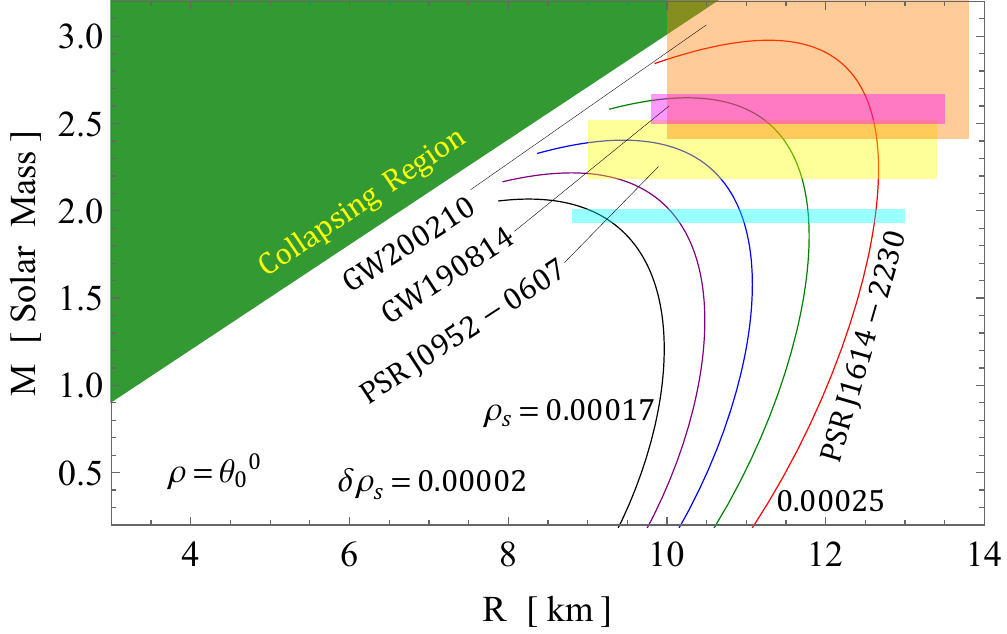}~~~~~
\includegraphics[height=6.2cm,width=8cm]{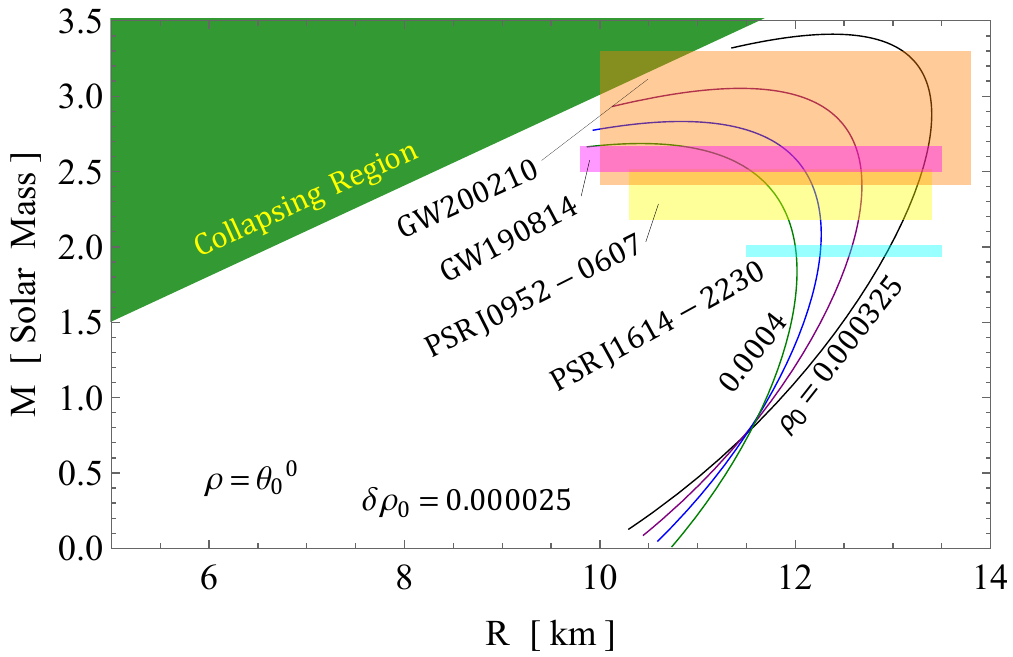}
\caption{$M-R$ curves for different values of $\sigma$ and $\rho_{_s}$ for $\rho$ mimicking $\theta_0^0$ for $\alpha = 100,~ \beta = 0.15,~ \rho_{_0} =0.00038/km^2,~\sigma=0.2$ and $\alpha = 100,~ \beta = 0.15,~ \rho_{_s} =0.00024/km^2,~\sigma=0.2$.}
\label{f8}
\end{figure*}

\section{Physical behavior of $M-R$ profiles}\label{sec7}

The contributions from Capozziello and his team \citep{Astashenok:2020qds, Nashed:2021sji, Astashenok:2021peo} have greatly enhanced our understanding of anisotropic stars and their mass-radius relationships. One of the significant breakthroughs in the study of compact stars comes from the work of \cite{Astashenok:2021peo}, which focused on the $\mathcal{R}^2$ model. They found that the mass-radius ($M-R$) curve in this context aligns with the general relativistic limit, around $3 M_{\odot}$. Notably, their research showed that the causal mass upper limit closely matches the maximum mass predicted by general relativity, placing it within the identified mass gap. Their investigation also delved into strange stars within the framework of modified gravity, particularly in relation to the secondary component of the GW190814 event. In another important study, \cite{Astashenok:2020qds} explored the existence of supermassive compact stars, pinpointing masses between $2.2$ and $2.3 M_{\odot}$ and radii of about $11$ km, all located within the gravity model of the axion $\mathcal{R}^2$. Building on this work, \cite{Astashenok:2020cqq} examined rotating neutron star models using the GM1 EOS, comparing their findings to static stars and presenting insights into the relationships between mass and central density. Their research identified realistic supermassive neutron stars with masses of $2.2$ to $2.3 M_\odot$ and a radius of $11$ km based on the APR EOS \citep{Astashenok:2020cfv}. In the context of perturbative $f(\mathcal{R})$ gravity, stable neutron stars were studied using FPS and SLy EOS, leading to a minimum radius of $9$ km and a maximum mass of $1.9 M_\odot$ \citep{Astashenok:2013vza}. Furthermore, they explored extreme neutron stars within the $f(G)$ and $f(\mathcal{R})$ theories, demonstrating the possibility of stars exceeding $4 M_\odot$ with radii between $12$ and $15$ km \citep{Astashenok:2014nua}. Their work also examined various EOS under different inflationary scenarios, highlighting that MPA1 is the only EOS that meets all the necessary constraints, allowing for masses greater than $2.5 M_\odot$ but below the $3 M_\odot$ causal limit \citep{Astashenok:2017dpo, Odintsov:2023ypt}.

\begin{table*}[!htp]
\footnotesize
\centering
\caption{The predicted radii of few high mass CSs corresponding to Fig. \ref{f6}.}\label{tab1}
\scalebox{0.7}{\begin{tabular}{|*{12}{c|} }
\hline
& & \multicolumn{10}{c|}{{Predicted $R$ (km) in $P_r=\theta_1^1$}} \\[0.15cm]
\cline{3-12}
{Objects} & {$\frac{M}{M_\odot}$} & \multicolumn{5}{c|}{ $\sigma$} & \multicolumn{5}{c|}{$\rho_{_0} \times 10^{-5}$} \\[0.15cm]
\cline{3-7} \cline{8-12}
&  & 0 & $0.05$ & $0.10$ & $0.15$ & $0.2$ & $35$ & $35$ & $37$ & $38$ & $39$ \\[0.15cm] \hline
PSR J1614-2230 \citep{Demorest:2010bx}  &  $1.97 \pm 0.04$  & $14.03_{-0.02}^{+0.03}$  &  $13.87_{-0.04}^{+0.02}$  &   $13.69_{-0.04}^{+0.05}$  &  $13.52_{-0.05}^{+0.04}$  &  $13.34_{-0.05}^{+0.06}$ & $14.70_{-0.01}^{+0.03}$ & $14.40_{-0.03}^{+0.02}$  & $14.12_{-0.03}^{+0.03}$ & $13.87_{-0.04}^{+0.03}$  & $13.63_{-0.03}^{+0.04}$  \\[0.15cm]
\hline
PSR J0952-0607 \citep{Romani:2022jhd} & $2.35 \pm 0.17$  & $13.64_{-0.31}^{+0.22}$  &  $13.38_{-0.45}^{+0.27}$  &   $13.09_{-}^{+0.34}$  &  $12.71_{-}^{+0.50}$  &  $-$ & $14.48_{-0.13}^{+0.13}$ & $14.11_{-0.21}^{+0.15}$  & $13.74_{-0.30}^{+0.20}$ & $13.39_{-0.45}^{+0.26}$  & $13.02_{-}^{+0.35}$  \\ [0.15cm]
\hline
GW190814 \citep{LIGOScientific:2020zkf} & $2.5-2.67$ & $13.20_{-0.40}^{+0.18}$ &  $12.67_{-}^{+0.33}$  & $-$ &  $-$ & $-$ & $14.28_{-0.12}^{+0.09}$  & $13.81_{-0.20}^{+0.12}$ & $13.30_{-0.40}^{+0.18}$ & $12.60_{-}^{+0.26}$  & $-$  \\[0.15cm]
\hline
GW200210 \citep{KAGRA:2021vkt} & $2.83^{+0.47}_{-0.42}$ &  $-$  &  $-$  & $-$  & $-$ &  $-$ & $13.86_{-}^{+0.58}$  &  $12.86_{-}^{+1.17}$  & $-$  & $-$ &  $-$ \\[0.15cm]
\hline
\end{tabular} }
\end{table*}

\begin{table*}[!htp]
\footnotesize
\centering
\caption{The predicted radii of few high mass CSs corresponding to Fig. \ref{f7}.}\label{tab2}
\scalebox{0.7}{\begin{tabular}{|*{12}{c|} }
\hline
& & \multicolumn{5}{c|}{{Predicted $R$ (km) in $P_r=\theta_1^1$}} & \multicolumn{5}{c|}{{Predicted $R$ (km) in $\rho=\theta_0^0$}} \\[0.15cm]
\cline{3-12}
{Objects} & {$\frac{M}{M_\odot}$} & \multicolumn{5}{c|}{ $\rho_s \times 10^{-5}$} & \multicolumn{5}{c|}{$\sigma$} \\[0.15cm]
\cline{3-7} \cline{8-12}
&  & 21 & $22$ & $23$ & $24$ & $25$ & $0$ & $0.05$ & $0.10$ & $0.15$ & $0.20$ \\[0.15cm] \hline
PSR J1614-2230 \citep{Demorest:2010bx}  &  $1.97 \pm 0.04$  & $-$  &  $12.33_{-0.13}^{+0.13}$  &   $13.17_{-0.07}^{+0.05}$  &  $13.87_{-0.04}^{+0.03}$  &  $14.54_{-0.02}^{+0.02}$ & $11.14_{-0.01}^{+0.01}$ & $11.42_{-0.01}^{+0.01}$  & $11.68_{-0.01}^{+0.01}$ & $11.95_{-0.01}^{+0.01}$  & $12.20_{-0.02}^{+0.01}$  \\[0.15cm]
\hline
PSR J0952-0607 \citep{Romani:2022jhd} & $2.35 \pm 0.17$  & $-$  &  $-$ & $-$  &  $13.38_{-0.45}^{+0.26}$  &  $14.26_{-0.18}^{+0.14}$ & $10.92_{-0.42}^{+0.17}$ & $11.24_{-0.30}^{+0.13}$  & $11.55_{-0.24}^{+0.11}$ & $11.84_{-0.19}^{+0.09}$  & $12.12_{-0.15}^{+0.07}$  \\ [0.15cm]
\hline
GW190814 \citep{LIGOScientific:2020zkf} & $2.5-2.67$ & $-$  &  $-$  &   $-$  &  $12.65_{-}^{+0.35}$  &  $14.00_{-0.16}^{+0.01}$ & $-$ & $10.74_{-}^{+0.24}$  & $11.19_{-0.51}^{+0.16}$ & $11.55_{-0.28}^{+0.13}$  & $11.89_{-0.19}^{+0.10}$ \\[0.15cm]
\hline
GW200210 \citep{KAGRA:2021vkt} & $2.83^{+0.47}_{-0.42}$ &  $-$  &  $-$  &   $-$  &  $-$  &  $13.42_{-}^{+0.78}$ & $-$ & $-$  & $-$ & $-$  & $-$ \\[0.15cm]
\hline
\end{tabular} }
\end{table*}
\begin{table*}[!htp]
\footnotesize
\centering
\caption{The predicted radii of few high mass CSs corresponding to Fig. \ref{f8}.}\label{tab3}
\scalebox{0.7}{\begin{tabular}{|*{11}{c|} }
\hline
& & \multicolumn{9}{c|}{{Predicted $R$ (km) in $\rho=\theta_0^0$}} \\[0.15cm]
\cline{3-11}
{Objects} & {$\frac{M}{M_\odot}$} & \multicolumn{4}{c|}{ $\rho_{_0} \times 10^{-4}$} & \multicolumn{5}{c|}{$\rho_{_s} \times 10^{-5}$} \\[0.15cm]
\cline{3-7} \cline{8-11}
&  & 3.25 & $3.50$ & $3.75$ & $4.00$ & $17$ & $19$ & $21$ & $23$ & $25$  \\[0.15cm] \hline
PSR J1614-2230 \citep{Demorest:2010bx}  &  $1.97 \pm 0.04$  & $12.94_{-0.03}^{+0.04}$  &  $13.56_{-0.01}^{+0.02}$  &   $12.26_{-0.01}^{+0.01}$  &  $12.01_{-0.01}^{+0.01}$  &  $9.18_{-0.18}^{+0.14}$ & $10.10_{-0.08}^{+0.06}$ & $10.94_{-0.03}^{+0.03}$  & $11.79_{-0.02}^{+0.01}$ & $12.62_{-0.01}^{+0.01}$   \\[0.15cm]
\hline
PSR J0952-0607 \citep{Romani:2022jhd} & $2.35 \pm 0.17$  & $13.22_{-0.09}^{+0.11}$  &  $12.68_{-0.03}^{+0.01}$  &   $12.21_{-0.13}^{+0.05}$  &  $11.79_{-0.28}^{+0.14}$  &  $-$ & $-$ & $10.12_{-}^{+0.56}$  & $11.55_{-0.31}^{+0.16}$ & $12.66_{-0.05}^{+0.01}$    \\ [0.15cm]
\hline
GW190814 \citep{LIGOScientific:2020zkf} & $2.5-2.67$ & $13.33_{-0.03}^{+0.03}$  &  $12.66_{-0.04}^{+0.01}$  &   $12.02_{-0.17}^{+0.09}$  &  $11.36_{-0.49}^{+0.20}$  &  $-$ & $-$ & $-$  & $11.05_{-}^{+0.23}$ & $12.57_{-0.08}^{+0.04}$   \\[0.15cm]
\hline
GW200210 \citep{KAGRA:2021vkt} & $2.83^{+0.47}_{-0.42}$ &  $13.39_{-0.06}^{+0.01}$  &  $12.40_{-}^{+0.28}$  &   $-$  &  $-$  &  $-$ & $-$ & $-$  & $-$ & $12.23_{-}^{+0.41}$   \\[0.15cm]
\hline
\end{tabular} }
\end{table*}

The $M-R$ curves we look at are based on the idea that the way matter is distributed in these fascinating stars can be described using a generalized polytropic EOS. To do this, we use a modified density model suggested by Mak \& Harko~\cite{Harko:2002prx}, along with the Tolman-Oppenheimer-Volkoff (TOV) equation. This approach allows us to explore how mass and radius relate for pulsars in both the GR and their extended MGD context. In our analysis, we have estimated the radii of four notable massive stellar objects: PSR J1614-2230 \citep{Demorest:2010bx}, PSR J0952-0607 \citep{Romani:2022jhd}, GW190814 \citep{LIGOScientific:2020zkf}, and GW200210 \citep{KAGRA:2021vkt}, using the $M-R$ curves that match their observed masses. Beyond determining maximum mass limits indicated by these curves, our findings open intriguing discussions about the nature of matter under extreme conditions. The differences we see in the $M-R$ relationships can give us valuable insight into the internal structure and stability of these stars. This exploration not only enhances our understanding of neutron stars, but also sheds light on their role in the universe, including the formation of gravitational waves and how matter behaves under incredible densities. In deriving the results presented above, we based our findings on three key assumptions that were somewhat arbitrary. Now, we want to explore how these assumptions affect the $M-R$ relationships we have established. By adjusting the parameters $\sigma$, $\rho_0$, and $\rho_s$, we can create more flexibility in how the EOS behaves at different densities. To do this, we generated several $M-R$ curves using randomly chosen values for these parameters, making sure that they still meet the necessary smoothness and subluminality criteria. Looking closely at the $M-R$ curves helps us uncover important insights about how changes in these three key parameters affect the properties of the solutions we have developed. This analysis gives us a better understanding of the underlying physical phenomena. We can see the variations in the $M-R$ curves illustrated in Figs. \ref{f6}, \ref{f7}, and \ref{f8}, which show how different values of $\sigma$, $\rho_{0}$, and $\rho_{s}$ influence the curves, with $P_r$ representing $\theta_1^1$ and $\rho$ representing $\theta_0^0$.

In Solution \ref{solA}, we investigate the behavior of the maximum mass under varying parameters, specifically focusing on $P_r$ which mimics $\theta_1^1$. The fixed parameters are $\alpha = 100$ and $\beta = 0.15$. With $\rho_{_0} = 0.00038/km^2$ and $\rho_{_s} = 0.00024/km^2$ constant, we observe that as $\sigma$ increases from 0 to 0.2, there is a significant decrease in the maximum mass. This maximum mass is highest at $\sigma=0$, as shown in the left panel of Fig. \ref{f6}. This trend suggests that lower values of $\sigma$ favor higher maximum masses, indicating an inverse relationship between $\sigma$ and maximum mass. In a separate analysis, where we fix $\sigma = 0.05$ and $\rho_{_s} = 0.00024/km^2$, we vary $\rho_{_0}$ from 0.00035$/km^2$ to 0.00039$/km^2$. Here, we again see a significant decrease in the maximum mass, which peaks at $\rho_{_0} = 0.00035/km^2$. The right panel of Fig. \ref{f6} illustrates this behavior, with a noticeable trend: as $\rho_{_0}$ increases, the maximum mass diminishes. This parallels the effect observed when varying $\sigma$, although the maximum mass at $\rho_{_0} = 0.00039/km^2$ is lower than at $\sigma = 0$, highlighting that both parameters exert similar influences on the maximum mass. Interestingly, when we again fix $\sigma = 0.05$ and $\rho_{_0} = 0.00038/km^2$, but vary $\rho_{_s}$ from 0.00021$/km^2$ to 0.00025$/km^2$, we observe a different trend where the maximum mass increases, peaking at $\rho_{_s} = 0.00025/km^2$. This behavior is illustrated in the left panel of Fig. \ref{f7}. The increase in maximum mass with $\rho_{_s}$ suggests that there is a range in which higher values of $\rho_{_s}$ positively influence maximum mass. The left panel of Fig. \ref{f7} shows how the increased range of $\rho_{_s}$ from 0.00021$/km^2$ to 0.00025$/km^2$ captures the combined effects of both $\rho_{_0}$ and $\sigma$. This comprehensive analysis indicates that varying these parameters not only affects the maximum mass but also highlights the nuanced interplay between them, reinforcing the importance of careful parameter selection in modeling.

In solution \ref{solB}, we explore the dynamics of maximum mass with respect to the parameter $\rho$, which mimics $\theta_0^0$, while keeping the fixed parameters at $\alpha = 100$ and $\beta = 0.15$. For this analysis, we set $\rho_{0} = 0.00038/km^2$ and $\rho_{s} = 0.00024/km^2$. As we increase the parameter $\sigma$ from 0 to 0.2, we observe a significant increase in maximum mass, which peaks at $\sigma = 0.2$. This trend is illustrated in the right panel of Fig. \ref{f7}. The results indicate that higher values of $\sigma$ are associated with higher maximum masses, suggesting a positive correlation between $\sigma$ and maximum mass in this context. In another scenario within Solution \ref{solB}, we keep $\sigma$ fixed at 0.05 and $\rho_{0} = 0.00038/km^2$, while varying $\rho_{s}$ from 0.00017$/km^2$ to 0.00025$/km^2$. Here, we again see a substantial increase in maximum mass, which peaks at $\rho_{s} = 0.00025/km^2$. This behavior is shown in the left panel of Fig. \ref{f8}. In particular, the maximum mass achieved at $\rho_{s} = 0.00025/km^2$ exceeds the maximum masses observed at higher values of $\sigma$, indicating that the influence of $\rho_{s}$ in this range is particularly strong. Additionally, we explore the effects of varying $\rho_{0}$ while fixing $\sigma = 0.05$ and $\rho_{s} = 0.00024/km^2$. As $\rho_{0}$ increases from 0.0000325$/km^2$ to 0.0004$/km^2$, we find that the maximum mass decreases significantly, reaching its peak at $\rho_{0} = 0.0000325/km^2$. This trend is presented in the right panel of Fig. \ref{f8}. The observed decrease in maximum mass with increasing $\rho_{0}$ aligns with previous findings regarding the effects of $\sigma$ and $\rho_{s}$, demonstrating that lower values of $\rho_{0}$ produce higher maximum masses. In general, the analysis indicates that while increasing $\sigma$ and $\rho_{s}$ increases the maximum mass, an increase in $\rho_{0}$ has the opposite effect. The relationship between these parameters highlights the complex interplay in the system, where varying $\rho_{s}$ and $\sigma$ leads to higher maximum masses, while increasing $\rho_{0}$ counteracts this trend. This multifaceted behavior underscores the importance of parameter selection in achieving desired outcomes within the model.

In summary, these findings illustrate the sensitivity of the maximum mass to changes in the parameters, revealing distinct patterns based on which parameters are manipulated in solutions \ref{solA} and \ref{solB}. Our study is based on the mass–radius relationships identified by various astrophysical groups, and we found that our results resonate well with four remarkable massive stellar objects: PSR J1614-2230 \citep{Demorest:2010bx}, PSR J0952-0607 \citep{Romani:2022jhd}, GW190814 \citep{LIGOScientific:2020zkf}, and GW200210 \citep{KAGRA:2021vkt}. The horizontal bands in our analysis represent the observational constraints derived from these objects, which help us to understand where our findings fit within the broader context of stellar physics. In this research, we have carefully fitted the data for these massive neutron stars, and interestingly, our approach suggests that there are many more potential candidates that could also be accommodated within these constraints. This indicates that the relationships between mass and radius for neutron stars are likely broader than just the four objects we have highlighted. Additionally, we have included several tables that provide predicted radii for high-mass compact stars that relate to the figures in our analysis. Specifically, Table \ref{tab1} presents the predicted radii for a selection of high-mass compact stars associated with Fig. \ref{f6}. Table \ref{tab2} details the predicted radii for those linked to Fig. \ref{f7}, while Table \ref{tab3} covers the predicted radii for high-mass compact stars relevant to Fig. \ref{f8}. These tables not only illustrate the range of possible stellar configurations but also reinforce the reliability of our mass–radius relationship findings.

So, when we try to understand what happens inside a NS, we look at how its mass and size are connected to the super-dense material it contains. To do that, we use a bunch of different ``recipes'', which scientists call models. These models are like our best educated guesses based on physics. Now, here's where it gets interesting: one of the biggest differences between these models is how the star's size changes as its mass increases. For some models, like SLy4 \citep{Chabanat:1997un}, SLy9 \citep{Chabanat:1997un}, SFHo \citep{Steiner:2012xt}, FSU \citep{Chen:2014sca}, and FSU2 \citep{Chen:2014sca}, if we make the star heavier than our sun, it always gets smaller -- like squeezing a balloon! (see Refs. \citep{Douchin:2001sv, Fortin:2016hny}) But other models, like DD2 \citep{Typel:2009sy}, DDME2 \citep{Lalazissis:2005de}, NL3$\omega \rho$ \citep{Horowitz:2000xj, Pais:2016xiu}, FSU2H \citep{Tolos:2016hhl}, and BigApple \citep{Fattoyev:2020cws}, do this weird thing where the star actually gets bigger for a while as we add mass! It's like the star is fighting back against gravity. Then we have models like NL3 \citep{Lalazissis:1996rd} and IUFSU \citep{Fattoyev:2010mx} that are just super sensitive -- a tiny change in mass makes a huge change in size. The cool thing is, even if a model starts out with the star getting bigger, eventually, as we keep adding mass, it has to start shrinking again before it goes kaboom and collapses into a BH. So, what is the reason these models act differently? Basically, it's because of the kind of nuclear forces they use inside. Like, even though the NL3 and NL3$\omega \rho$ models are exactly the same when it comes to the basic ``stuff'' inside (what they call ``symmetric nuclear matter''), they handle another ingredient, ``symmetry energy'', in different ways. The NL3$\omega \rho$ model has less of this ``symmetry energy'' and it doesn't change as quickly at a certain point (``flatter slope at saturation'') compared to NL3. This difference makes NL3$\omega \rho$ stars, especially the smaller and medium-sized ones, have smaller sizes than NL3 stars. That is why the mass-radius relationship does that weird ``backbending'' thing. 

We see something similar when we compare the FSU2 and FSU2H models. They are not perfectly the same in terms of that basic ``stuff'' (``symmetric nuclear matter properties''), but they're pretty similar. However, the BigApple model, like FSU2H, has a kinda small symmetry energy and slope at saturation, but its EOS is seriously stiff when things get really dense. This means the model can handle a 2.6 $M_{\odot}$, so it figures that low- and middle-mass stars are tiny, but the supermassive ones are huge. FSU and FSU2, both of them, have a soft isoscalar thing going on at all densities, and that's totally why their $M-R$ curves are always sloping downwards for stars bigger than the sun. The SFHo EOS was made to create small-radius stars, which it does by being soft when things get dense and having a soft symmetry energy. SLy4 and SLy9 act just like SFHo. EOSs that predict $M-R$ curves that are basically straight up and down for most stars have properties somewhere in the middle: NL3 has a strong symmetry energy and a strong nuclear matter EOS, while IUFSU is based on FSU, which makes it stiffer when things get dense. 

The way the $M-R$ curve behaves tells us about the hadronic interaction, with the backbending usually meaning the stuff inside is getting harder to squeeze. For the models with a positive or infinite slope, the curve will always flip to a negative slope for really big stars before they hit their max mass. This flip happens because the strong force and gravity are fighting it out. When new particles show up, like hyperons or quarks breaking free, it can make this flip happen sooner -- check out Fig. 4 of \cite{Fortin:2016hny} and Fig. 4 of \cite{Ferreira:2020kvu}. Also, making the EOS stiffer when things get dense has been linked to the start of a quarkyonic phase \cite{McLerran:2018hbz, Zhao:2020dvu}. Unlike quarks breaking free or hyperons showing up, this phase actually causes the backbending in the $M-R$ curve--see Fig. 4 of Ref. \citep{McLerran:2018hbz}. As a result, our models for $M-R$ curves -- using solution \ref{solA}, which focuses on $P_r$ (kind of like $\theta_1^1$), and solution \ref{solB}, which focuses on $\rho$ (kind of like $\theta_0^0$) -- employing a generalized polytropic EOS are well-matched with the microscopic phenomenological models described above. These models determine the $M-R$ curves that link astrophysical observations of NSs to the EOS of baryon matter. While the symmetric nuclear matter properties derived from these solutions are not exactly the same, they are quite similar.

\section{Concluding remarks}\label{sec8} 

In our present investigation, we have attempted to explore the idea that the matter distribution in the stars can be described by using a suitable generalized polytropic equation of state. To achieve this target, we (i) use a modified energy-density profile proposed by Mak \& Harko~\cite{Harko:2002prx} alongside the TOV equation and (ii) investigate the relationship between the masses and the radii for pulsars in both GR and the extended framework of the modified geometric distribution (MGD) method. After these operations, we estimate the radii of four notable massive stellar objects: PSR J1614-2230, PSR J0952-0607, GW190814, and GW200210 via the mass-radius ($M-R$) curves that align with their observed masses. The obtained results are profound and physically viable as far as the configuration of strange quark stars is concerned.

Some salient features of the study are as follows:
\begin{itemize}
  \item {\bf Physical behavior of energy density and pressure profiles}: 

(i) The effective energy density, which we refer to as $\rho^{\text{eff}}$ and have illustrated in Fig. \ref{f1}, exhibits an interesting pattern in connection with deformed SS models. Here we note that at the center of these stars, the energy density is at its highest point, whereas toward the surface the density gradually decreases, reaching its lowest value right at the surface. One can observe the effect of the decoupling parameter $\sigma$ which has a definite impact on the effective energy density ($\rho^{\text{eff}}$) throughout the entire range of the radial coordinates $r$, as is evident from the different panels of Fig. \ref{f1} which we have thoroughly discussed in the relevant places in Subsection \ref{sec5.1}. 

(ii) In this context, we also note that for the first solution -- $P_r(r) = \theta^1_1(r)$~[\ref{solA}], the core density is relatively low, while the second solution -- $\rho(r) = \theta^0_0(r)$~[\ref{solB}] provides a slight increase in the core density. Obviously, this difference of attributes is mainly due to the decoupling effect, as has been explained in Subsection \ref{sec5.1} how the decoupling effect becomes influential to the distribution of density in stellar models.

(iii) In a similar way, one can check the effective radial ($P_r^{\text{eff}}$) and tangential ($P_t^{\text{eff}}$) pressures which reveals some amazing differences between the two solutions. We have highlighted all these patterns to show how the decoupling effect $\sigma$ influences the fluid particles within the star. 
  
  \item {\bf Physical behavior of anisotropic profiles}:
  
(i) From Fig. \ref{f3}, we get the clear cut information on the anisotropy parameter, $\Delta^{\text{eff}}$, which starts at zero at the center of the star and gradually increases as one moves toward the outer boundary. In the case of the first solution~\ref{solA}, i.e., $P_r(r) = \theta^1_1(r)$, the anisotropy parameter reaches its highest point just before touching the surface. This upward trend is an indication that the effective radial stresses are stronger than the transverse stresses, thus resulting in an outward repulsive force.

(ii) In the second solution \ref{solB}, i.e. $\rho(r) = \theta^0_0(r)$, the parameter $\sigma$ increases and makes a sharp surge in anisotropy, which is involved in several intricate physical processes within the stellar configuration. It should be remembered that phase transitions can change the form of the EOS and thus can affect the balance between radial and tangential stresses. Moreover, dissipative effects (e.g. viscosity and heat conductivity) can also play a vital key role in creating anisotropic stresses. Therefore, inside the stellar configuration, we are getting an environment of dynamics that greatly enhances its anisotropies and certainly has its own complex interactions toward its stability and evolution over time.

  \item {\bf Stability analysis via adiabatic index}: 

To achieve the stability of anisotropic stellar configurations we check the status of the adiabatic index $\Gamma$, which for isotropic fluids in a Newtonian framework needs to be greater than $\frac{4}{3}$. However, in the context of GR, the critical value becomes $\Gamma_{cr} = \frac{4}{3} + \frac{19}{42} \frac{2M}{R}$, where $\mathcal{U} = \frac{2M}{R}$ represents the compactness parameter.

In the above situation, the behavior of the adiabatic index $\Gamma$ can be exhibited in Fig. \ref{f4} where we have employed the same sets of parameters as in the case of Figs. \ref{f1} to \ref{f3}. Interesting features from Fig. \ref{f4} are as follows: (i) one can notice a linkage between the adiabatic index and the stability of the stellar structure, (ii) the first solution relates to the pressure inside the star as can be described via $P_r(r) = \theta^1_1(r)$~[\ref{solA}] which indicates the pressure-dependent stability, and (iii) the second solution focuses on the density of the star via $\epsilon(r) = \theta^0_0(r)$~[\ref{solA}] which indicates the density distribution in connection with stability. In conclusion, all of these results shed light on how both pressure and density play important roles in maintaining stellar stability.

 \item {\bf Stability analysis via cracking criterion}: 

Following the cracking concept of \cite{Herrera1992} we have depicted in Fig. \ref{f5} the principle of causality. In this graphical plot, both the sound speeds ($v^{2}_{r}$ and $v^{2}_{t}$) stay below the speed of sound, i.e., below unity, as such across all increasing values of $\sigma$ in both solution~\ref{solA} (left panel) and solution~\ref{solB} (right panel). This result is a strong indication that the impacts of MGD within GR remain favorable and maintain the causality principle. 

\item {\bf Stability analysis via Harrison-Zel’dovich-Novikov criteria}: 

To explore a path to understand the connection between the stellar mass and the central energy density $\rho^{\text{eff}}_0$, we have drawn Fig. \ref{fig5a}. From the graphical graph, it is clear that $M(\rho^{\text{eff}}_0)$ is positive and increasing and therefore indicates that $\frac{dM}{d\rho^{\text{eff}}_0}$ is positive. This explicitly expresses the stability of the stellar configuration. 

 \item {\bf Physical behavior of $M-R$ profiles}: 

The mass–radius relationship as we have discussed in a comprehensive manner in Section \ref{sec7}, can be understood by (i) the table \ref{tab1} which presents the predicted radii for a selection of high-mass compact stars associated with Fig. \ref{f6}, (ii) the table \ref{tab2} which details the predicted radii for those linked to Fig.~\ref{f7} and (iii) the Table \ref{tab3} which covers the predicted radii for high-mass compact stars relevant to Fig.~\ref{f8}. The basic idea behind the $M-R$ curves is to comprehend and visualize the matter distribution within the stellar configuration via the adopted generalized polytropic EOS. We have described the related methodology where we use a modified density model suggested by~\cite{Harko:2002prx}, along with the TOV equation in the framework of GR and their extended MGD technique. In the analysis, using the $M-R$ curves that match their observed masses, we have estimated the radii of four notable massive stellar objects, viz. PSR J1614-2230 \citep{Demorest:2010bx}, PSR J0952-0607 \citep{Romani:2022jhd}, GW190814 \citep{LIGOScientific:2020zkf}, and GW200210 \citep{KAGRA:2021vkt}. Thus we have determined maximum mass limits as indicated by these curves, and also we open up intriguing discussions about the nature of matter in extreme conditions, insights into the internal structure and stability of these stars. The variations in the $M-R$ curves as exhibited in Figs. \ref{f6}, \ref{f7} and \ref{f8} clearly indicate how different values of $\sigma$, $\rho_{0}$, and $\rho_{s}$ influence the curves based on the solution \ref{solA} and \ref{solB}.

\end{itemize}

In the present work, we have investigated the matter distribution in stellar systems using a modified energy density profile proposed by Harko et al. \citep{Harko:2002prx}. However, there may be ample scopes to check the issue based on other types of profiles as available in the literature, such as the Navarro-Frenk-White (NFW) profile \citep{NFS1,NFS2}, Einasto profile \citep{Merritt1,Merritt2}, three-parameter dark halo model \citep{Begeman}, Universal
Rotation Curve (URC) profile \citep{URC1,URC2,URC3}, dilute Bose-Einstein condensate, and Scalar Field Dark Matter (SFDM) profile \citep{MauryaPDU}. Some of these profiles already have shown their robust effect on stellar as well as galactic systems in the presence of dark matter (DM) \citep{URC1,URC2,URC3}. Therefore, one may be interested in applying some of these profiles in connection with the issue of the present investigation to verify physical features of the stellar configuration with and without DM. 

 \section*{Acknowledgments}
The author S. K. Maurya acknowledges that the Ministry of Higher Education, Research, and Innovation (MoHERI) supported this research work through the project  BFP/RGP/CBS/24/203. SKM also thankful for continuous support and encouragement from the administration of the University of Nizwa for this research work. AE thanks the National Research Foundation of South Africa for the award of a postdoctoral fellowship. KNS and SR are also thankful to the authorities of the Inter-University Centre for Astronomy and Astrophysics, Pune, India for providing the research facilities where SR is specifically expresses thanks to ICARD of IUCAA at GLA University.

\appendix
\section{The expressions for coefficients are given as follows:  }
\begin{small}
\begin{eqnarray}
    &&\hspace{-1.2cm} \mathcal{G}_{0}(r)=\Big[16 \pi  \big(15 R^2-24 \rho_s \pi  r^4  +8 \rho_0 \pi  \left(3 r^2-5 R^2\right) r^2\big) \big[\rho_s^2 \alpha  r^4+\rho_s R^2 \beta  r^2 +\rho_0^2  \left(r^2-R^2\right)^2 \alpha  \nonumber\\&&\hspace{0.1cm} -\rho_0 \left(r^2-R^2\right) \left(2 \rho_s \alpha  r^2+R^2 \beta \right) +R^4 \gamma \big] \big[3 (\rho_0-\rho_s)^2 \alpha  r^4-2 (\rho_0-\rho_s) R^2 (2 \rho_0 \alpha +\beta ) r^2\nonumber\\&&\hspace{0.1cm}+R^4  (\alpha  \rho_0^2+\beta  \rho_0+\gamma )\big] r^2-2 (\rho_0-\rho_s)  \left(-24 \rho_s \pi  r^4+8 \rho_0 \pi  \left(3 r^2-5 R^2\right) r^2+15 R^2\right)  \big(-2 \rho_s \alpha  r^2\nonumber\\&&\hspace{0.1cm}+2 \rho_0 \left(r^2-R^2\right) \alpha -R^2 \beta \big)  \big[8 (\rho_0-\rho_s)^2 \pi  \alpha  r^6-8 (\rho_0-\rho_s) \pi  R^2 (2 \rho_0 \alpha +\beta ) r^4+R^4 (8 \rho_0^2 \pi  \alpha  r^2\nonumber\\&&\hspace{0.1cm} +8 \rho_0 \pi  \beta  r^2+8 \pi  \gamma  r^2+1)\big] r^2-16 \pi (6 \rho_0 r^2-6 \rho_s r^2-5 \rho_0 R^2) \big[\rho_s^2 \alpha  r^4+\rho_s R^2 \beta  r^2+\rho_0^2 \left(r^2-R^2\right)^2 \alpha \nonumber\\&&\hspace{0.1cm} -\rho_0 \left(r^2-R^2\right) (2 \rho_s \alpha  r^2+R^2 \beta )+R^4 \gamma \big] \Big(8 (\rho_0-\rho_s)^2 \pi  \alpha  r^6-8 (\rho_0-\rho_s) \pi  R^2 (2 \rho_0 \alpha +\beta ) r^4\nonumber\\&&\hspace{0.1cm}+R^4 \big(8 \rho_0^2 \pi  \alpha  r^2+8 \rho_0 \pi  \beta  r^2+8 \pi  \gamma  r^2+1\big)\Big) r^2-2 \big[-24 \rho_s \pi  r^4+8 \rho_0 \pi  \left(3 r^2-5 R^2\right) r^2+15 R^2\big] \nonumber\\&&\hspace{0.1cm} \times \Big(\rho_s^2 \alpha  r^4+\rho_s R^2 \beta  r^2  +\rho_0^2 \left(r^2-R^2\right)^2 \alpha -\rho_0 \left(r^2-R^2\right) \left(2 \rho_s \alpha  r^2+R^2 \beta \right)+R^4 \gamma \Big) \Big\{8 (\rho_0-\rho_s)^2 \pi  \nonumber\\&&\hspace{0.1cm} \times \alpha  r^6-8 (\rho_0-\rho_s) \pi  R^2 (2 \rho_0 \alpha +\beta ) r^4+R^4 \left(8 \rho_0^2 \pi  \alpha  r^2+8 \rho_0 \pi  \beta  r^2+8 \pi  \gamma  r^2+1\right)\Big\}\Big],\nonumber\\ 
    \end{eqnarray}
    \end{small}
\begin{small}
\begin{eqnarray}
  &&\hspace{-1.2cm}  \mathcal{F}_1(r)={8 (\rho_0-\rho_s)^2 \pi  R^2 \alpha  r^6-8 (\rho_0-\rho_s) \pi  R^4 (2 \rho_0 \alpha +\beta ) r^4+R^6 \big(8 \rho_0^2 \pi  \alpha  r^2+8 \rho_0 \pi  \beta  r^2+8 \pi  \gamma  r^2+1\big)},\nonumber\\
  &&\hspace{-1.2cm} \mathcal{F}_2(r)=\big(8 (\rho_0-\rho_s)^2 \pi  R \alpha  r^6-8 (\rho_0-\rho_s) \pi  R^3 (2 \rho_0 \alpha +\beta ) r^4+R^5 \big(8 \rho_0^2 \pi  \alpha  r^2+8 \rho_0 \pi  \beta  r^2+8 \pi  \gamma  r^2+1\big)\big)^2,\nonumber\\
  &&\hspace{-1.2cm} \mathcal{F}_3(r)=\big(8 (\rho_0-\rho_s)^2 \pi  \alpha  r^6-8 (\rho_0-\rho_s) \pi  R^2 (2 \rho_0 \alpha +\beta ) r^4+R^4 \big(8 \rho_0^2 \pi  \alpha  r^2+8 \rho_0 \pi  \beta  r^2+8 \pi  \gamma  r^2+1\big)\big)^2, \nonumber\\
 &&\hspace{-1.2cm} \mathcal{F}_4(r)=806400 \rho_s^6 \pi ^3 \big(r^2-R^2\big)^2 \alpha ^4 r^{18}+3840 \rho_s^5 \pi ^3 R^2 \alpha ^3 \big(63 (10 \beta -3) r^4+12 R^2 (31-105 \beta ) r^2+35 R^4 (18 \beta -5)\big) r^{16}\nonumber\\&&\hspace{-0.4cm}+1200 \rho_s^4 \pi ^2 R^4 \alpha ^2 \big(15 \big(39 r^4-70 R^2 r^2+31 R^4\big) \alpha +8 \pi  \big(9 \big(30 \beta ^2-14 \beta +20 \alpha  \gamma +1\big) r^6-10 R^2 \big(54 \beta ^2-25 \beta +36 \alpha  \gamma \nonumber\\&&\hspace{-0.4cm}+2\big) r^4+R^4 \big(270 \beta ^2-116 \beta +180 \alpha  \gamma +11\big) r^2\big)\big) r^{12}+480 \rho_s^3 \pi ^2 R^6 \alpha  \big(240 \pi  \big(10 \beta ^3-5 \beta ^2+30 \alpha  \gamma  \beta +\beta -5 \alpha  \gamma \big) r^6\nonumber\\&&\hspace{-0.4cm}+\big(-32 \pi  \beta  \big(150 \beta ^2-76 \beta +17\big) R^2-75 \alpha  \beta  \big(192 \pi  R^2 \gamma -35\big)+4 \alpha  \big(608 \pi  R^2 \gamma -45\big)\big) r^4+2 \big(8 \pi  \big(150 \beta ^3-69 \beta ^2\nonumber\\&&\hspace{-0.4cm}+(450 \alpha  \gamma +19) \beta -69 \alpha  \gamma \big) R^4+\alpha  (178-2325 \beta ) R^2\big) r^2+R^4 \alpha  (2025 \beta -164)\big) r^{10}+24 \rho_s^2 \pi  R^8 \big(160 \pi ^2 \big(9 \big(5 \beta ^4\nonumber\\&&\hspace{-0.4cm}-2 \beta ^3+(60 \alpha  \gamma +1) \beta ^2-12 \alpha  \gamma  \beta +2 \alpha  \gamma  (15 \alpha  \gamma +1)\big) r^4-3 R^2 \big(30 \beta ^4-13 \beta ^3+(360 \alpha  \gamma +7) \beta ^2-78 \alpha  \gamma  \beta +2 \alpha  \gamma \nonumber\\&&\hspace{-0.4cm}\times (90 \alpha  \gamma +7)\big) r^2+R^4 \big(45 \beta ^4-17 \beta ^3+12 (45 \alpha  \gamma +1) \beta ^2-102 \alpha  \gamma  \beta +6 \alpha  \gamma  (45 \alpha  \gamma +4)\big)\big) r^4+4 \pi  \alpha  \big(9 \big(775 \beta ^2-90 \beta \nonumber\\&&\hspace{-0.4cm}+775 \alpha  \gamma +11\big) r^4-3 R^2 \big(4050 \beta ^2-545 \beta +4050 \alpha  \gamma +77\big) r^2+R^4 \big(5175 \beta ^2-735 \beta +5175 \alpha  \gamma +128\big)\big) r^2+675 \nonumber\\&&\hspace{-0.4cm} \times \big(3 r^4-5 R^2 r^2+2 R^4\big) \alpha ^2\big) r^6+4 \rho_s \pi  R^{10} \big(8640 \pi ^2 \gamma  \big(10 \beta ^3-\beta ^2+30 \alpha  \gamma  \beta +\beta -\alpha  \gamma \big) r^8-12 \pi  \big(225 \big(64 \pi  R^2 \gamma -9\big) \beta ^3\nonumber\\&&\hspace{-0.4cm}+\big(270-2240 \pi  R^2 \gamma \big) \beta ^2-81 (150 \alpha  \gamma +1) \beta +160 \pi  R^2 \gamma  (270 \alpha  \gamma +11) \beta +20 \alpha  \gamma  \big(27-112 \pi  R^2 \gamma \big)\big) r^6+3 \big(8 \pi  \beta  \big(520 \pi  \nonumber\\&&\hspace{-0.4cm} \times \gamma  R^2+75 \beta ^2 \big(48 \pi  R^2 \gamma -23\big)+\beta  \big(290-440 \pi  R^2 \gamma \big)-101\big) R^2+5 \alpha  \big(-704 \pi ^2 \gamma ^2 R^4+928 \pi  \gamma  R^2+45 \beta  \big(384 \pi ^2 \gamma ^2 R^4\nonumber\\&&\hspace{-0.4cm}-368 \pi  \gamma  R^2+15\big)-81\big)\big) r^4+10 \big(2 \pi  R^4 \big(855 \beta ^3-150 \beta ^2+5130 \alpha  \gamma  \beta +71 \beta -300 \alpha  \gamma \big)-81 R^2 \alpha  (20 \beta -3)\big) r^2+45 R^4 \alpha \nonumber\\&&\hspace{-0.4cm} \times (135 \beta -23)\big) r^4+R^{12} \big(5760 \pi ^3 \gamma ^2 \big(30 \beta ^2+2 \beta +20 \alpha  \gamma +1\big) r^{10}-48 \pi ^2 \gamma  \big(320 \pi  \gamma  (15 \alpha  \gamma +1) R^2+75 \beta ^2 \big(96 \pi  R^2 \gamma -23\big)\nonumber\\&&\hspace{-0.4cm}-3 (575 \alpha  \gamma +7)+\beta  \big(90-80 \pi  R^2 \gamma \big)\big) r^8+120 \pi  \big(15 \big(96 \pi ^2 \gamma ^2 R^4-76 \pi  \gamma  R^2+3\big) \beta ^2+\big(100 \pi  R^2 \gamma -9\big) \beta +2 \gamma  \big(40 \pi ^2 \gamma \nonumber\\&&\hspace{-0.4cm} \times  (12 \alpha  \gamma +1) R^4-6 \pi  (95 \alpha  \gamma +2) R^2+45 \alpha \big)\big) r^6+5 \big(4 \pi  \big(100 \pi  \gamma  R^2+135 \beta ^2 \big(20 \pi  R^2 \gamma -3\big)-3 \beta  \big(80 \pi  R^2 \gamma -39\big)\big) R^2\nonumber\\&&\hspace{-0.4cm}+135 \alpha  \big(80 \pi ^2 \gamma ^2 R^4-24 \pi  \gamma  R^2+1\big)\big) r^4+900 \big(\pi  R^4 \big(3 \beta ^2-\beta +6 \alpha  \gamma \big)-R^2 \alpha \big) r^2+225 R^4 \alpha \big),\nonumber\\
 &&\hspace{-1.2cm} \mathcal{F}_5(r)=6720 \rho_s^3 \pi  \big(r^2-R^2\big)^2 \alpha ^3 r^8+16 \rho_s^2 \pi  R^2 \alpha ^2 \big(63 (10 \beta -3) r^4+3 R^2 (121-420 \beta ) r^2+2 R^4 (315 \beta -47)\big) r^6+2 \rho_s R^4 \alpha \nonumber\\&&\hspace{-0.4cm} \times \big(15 \big(39 r^4-58 R^2 r^2+19 R^4\big) \alpha +8 \pi  \big(9 \big(30 \beta ^2-14 \beta +20 \alpha  \gamma +1\big) r^6-R^2 \big(540 \beta ^2-247 \beta +360 \alpha  \gamma +23\big) r^4+R^4 \nonumber\\&&\hspace{-0.4cm} \times\big(270 \beta ^2-41 \beta +180 \alpha  \gamma +14\big) r^2\big)\big) r^2+R^6 \big(48 \pi  \big(10 \beta ^3-5 \beta ^2+30 \alpha  \gamma  \beta +\beta -5 \alpha  \gamma \big) r^6+\big(\alpha  \big(-2880 \pi  \beta  \gamma  R^2+496 \nonumber\\&&\hspace{-0.4cm} \times\pi  \gamma  R^2+525 \beta -36\big)-16 \pi  R^2 \beta  \big(60 \beta ^2-31 \beta +8\big)\big) r^4+10 \big(8 \pi  \beta  \big(6 \beta ^2+18 \alpha  \gamma +1\big) R^4+\alpha  (7-75 \beta ) R^2\big) r^2+5 R^4 \nonumber\\&&\hspace{-0.4cm} \times \alpha  (45 \beta -2)\big),\nonumber\\
 &&\hspace{-1.2cm} \mathcal{F}_6(r)=-675 \big(r^2-R^2\big)^2 \big(R^2-3 r^2\big) \alpha ^2 R^8+4 \pi  r^2 \alpha  \big(43875 \rho_s^2 \alpha ^2 r^{10}-225 \rho_s R^2 \alpha  (505 \rho_s \alpha -175 \beta +12) r^8+3 R^4 \big(31875 \rho_s^2 \alpha ^2\nonumber\\&&\hspace{-0.4cm} -5 \rho_s (6675 \beta -532) \alpha +3 \big(775 \beta ^2-90 \beta +775 \alpha  \gamma +11\big)\big) r^6-3 R^6 \big(8625 \rho_s^2 \alpha ^2-5 \rho_s (5475 \beta -452) \alpha +5775 \gamma  \alpha +5775 \beta ^2\nonumber\\&&\hspace{-0.4cm} -820 \beta +121\big) r^4+15 R^8 \big(915 \beta ^2-130 \beta -25 \rho_s \alpha  (57 \beta -4)+915 \alpha  \gamma +27\big) r^2-25 R^{10} \big(135 \beta ^2-12 \beta +135 \alpha  \gamma +5\big)\big) R^4\nonumber\\&&\hspace{-0.4cm}+480 \pi ^2 r^4 \big(r^2-R^2\big) \big(1050 \rho_s^4 \big(r^2-R^2\big)^2 \alpha ^4 r^8+10 \rho_s^3 R^2 \alpha ^3 \big(21 (10 \beta -3) r^4+2 R^2 (61-210 \beta ) r^2+R^4 (210 \beta -43)\big) r^6\nonumber\\&&\hspace{-0.4cm} +5 \rho_s^2 R^4 \alpha ^2 \big(9 \big(30 \beta ^2-14 \beta +20 \alpha  \gamma +1\big) r^4-2 R^2 \big(270 \beta ^2-124 \beta +180 \alpha  \gamma +11\big) r^2+R^4 \big(270 \beta ^2-74 \beta +180 \alpha  \gamma +13\big)\big) r^4\nonumber\\&&\hspace{-0.4cm} +2 \rho_s R^6 \alpha  \big(15 \big(10 \beta ^3-5 \beta ^2+30 \alpha  \gamma  \beta +\beta -5 \alpha  \gamma \big) r^4-2 R^2 \big(150 \beta ^3-77 \beta ^2+450 \alpha  \gamma  \beta +19 \beta -77 \alpha  \gamma \big) r^2+R^4 \big(150 \beta ^3\nonumber\\&&\hspace{-0.4cm} -31 \beta ^2+450 \alpha  \gamma  \beta +23 \beta -31 \alpha  \gamma \big)\big) r^2+R^8 \big(3 \big(5 \beta ^4-2 \beta ^3+(60 \alpha  \gamma +1) \beta ^2-12 \alpha  \gamma  \beta +2 \alpha  \gamma  (15 \alpha  \gamma +1)\big) r^4-2 R^2 \big(15 \beta ^4\nonumber\\&&\hspace{-0.4cm}-7 \beta ^3+4 (45 \alpha  \gamma +1) \beta ^2-42 \alpha  \gamma  \beta +2 \alpha  \gamma  (45 \alpha  \gamma +4)\big) r^2+5 R^4 \big(3 \beta ^4+(36 \alpha  \gamma +1) \beta ^2+2 \alpha  \gamma  (9 \alpha  \gamma +1)\big)\big)\big),\nonumber\\
&&\hspace{-1.2cm} \mathcal{F}_7(r)=420 \rho_s^2 \alpha ^2 r^8-42 \rho_s R^2 \alpha  (20 \rho_s \alpha -10 \beta +3) r^6+3 R^4 \big(140 \rho_s^2 \alpha ^2-40 \rho_s (7 \beta -2) \alpha +20 \gamma  \alpha +30 \beta ^2-14 \beta +1\big) r^4-2 R^6 \nonumber\\&&\hspace{-0.4cm} \times\big(90 \beta ^2-41 \beta +\rho_s \alpha  (17-210 \beta )+60 \alpha  \gamma +4\big) r^2+5 R^8 \big(18 \beta ^2+12 \alpha  \gamma +1\big),\nonumber\\
&&\hspace{-1.2cm}   \mathcal{F}_8(r)=3200 \rho_s^8 \pi ^3 \alpha ^4 r^{22}+1280 \rho_s^7 \pi ^3 R^2 \alpha ^3 (10 \beta -3) r^{20}+80 \rho_s^6 \pi ^2 R^4 \alpha ^2 \big(8 \pi  \big(30 \beta ^2-14 \beta +20 \alpha  \gamma +1\big) r^2+65 \alpha \big) r^{16}+80 \rho_s^5 \pi ^2 \nonumber\\&&\hspace{-0.4cm} \times R^6 \alpha  \big(16 \pi  \beta  \big(10 \beta ^2-5 \beta +1\big) r^2-4 \alpha  \big(20 \pi  \gamma  r^2+3\big)+5 \alpha  \beta  \big(96 \pi  \gamma  r^2+35\big)\big) r^{14}+4 \rho_s^4 \pi  R^8 \big(160 \pi ^2 \big(5 \beta ^4-2 \beta ^3+(60 \alpha  \gamma\nonumber\\&&\hspace{-0.4cm} +1) \beta ^2-12 \alpha  \gamma  \beta +2 \alpha  \gamma  (15 \alpha  \gamma +1)\big) r^4+4 \pi  \alpha  \big(775 \beta ^2-90 \beta +775 \alpha  \gamma +11\big) r^2+225 \alpha ^2\big) r^{10}+4 \rho_s^3 \pi  R^{10} \big(4 \pi  \beta  \big(80 \pi  \gamma  r^2\nonumber\\&&\hspace{-0.4cm} -10 \beta  \big(8 \pi  \gamma  r^2+3\big)+25 \beta ^2 \big(32 \pi  \gamma  r^2+9\big)+9\big) r^2-5 \alpha  \big(8 \pi  \gamma  r^2+3\big)^2+75 \alpha  \beta  \big(128 \pi ^2 \gamma ^2 r^4+72 \pi  \gamma  r^2+5\big)\big) r^8+\rho_s^2 R^{12} \nonumber\\&&\hspace{-0.4cm} \times\big(8 \pi  \big(2 \pi  \gamma  \big(40 \pi  \gamma  r^2+7\big) r^2+5 \beta  \big(32 \pi ^2 \gamma ^2 r^4-12 \pi  \gamma  r^2-3\big)+25 \beta ^2 \big(96 \pi ^2 \gamma ^2 r^4+46 \pi  \gamma  r^2+3\big)\big) r^2+25 \alpha  \big(512 \pi ^3 \gamma ^3 r^6\nonumber\\&&\hspace{-0.4cm} +368 \pi ^2 \gamma ^2 r^4+48 \pi  \gamma  r^2+3\big)\big) r^4+10 \rho_s R^{14} \big(2 \pi  \gamma  \big(64 \pi ^2 r^4 \gamma ^2-3\big) r^2+5 \beta  \big(256 \pi ^3 \gamma ^3 r^6+152 \pi ^2 \gamma ^2 r^4+18 \pi  \gamma  r^2+1\big)\big) r^2\nonumber\\&&\hspace{-0.4cm}+25 R^{16} \gamma  \big(128 \pi ^3 \gamma ^3 r^6+80 \pi ^2 \gamma ^2 r^4+12 \pi  \gamma  r^2+1\big),\nonumber
 \end{eqnarray}
 \newpage
\begin{eqnarray}
 &&\hspace{-1.2cm} \mathcal{F}_9(r)=76800 \rho_s^7 \pi ^3 \big(r^2-R^2\big) \alpha ^4 r^{20}+1280 \rho_s^6 \pi ^3 R^2 \alpha ^3 \big(21 (10 \beta -3) r^2+2 R^2 (31-105 \beta )\big) r^{18}+160 \rho_s^5 \pi ^2 R^4 \alpha ^2 \big(15 \big(39 r^2\nonumber\\&&\hspace{-0.4cm} -35 R^2\big) \alpha +8 \pi  \big(9 r^4 \big(30 \beta ^2-14 \beta +20 \alpha  \gamma +1\big)-5 r^2 R^2 \big(54 \beta ^2-25 \beta +36 \alpha  \gamma +2\big)\big)\big) r^{14}+80 \rho_s^4 \pi ^2 R^6 \alpha  \big(240 \pi  \big(10 \beta ^3\nonumber\\&&\hspace{-0.4cm}-5 \beta ^2+30 \alpha  \gamma  \beta +\beta -5 \alpha  \gamma \big) r^4+\big(-16 \pi  \beta  \big(150 \beta ^2-76 \beta +17\big) R^2-75 \alpha  \beta  \big(96 \pi  R^2 \gamma -35\big)+4 \alpha  \big(304 \pi  R^2 \gamma -45\big)\big) r^2\nonumber\\&&\hspace{-0.4cm} +R^2 \alpha  (178-2325 \beta )\big) r^{12}+8 \rho_s^3 \pi  R^8 \big(4 \pi  \alpha  \big(6 r^2 \big(775 \beta ^2-90 \beta +775 \alpha  \gamma +11\big)-R^2 \big(4050 \beta ^2-545 \beta +4050 \alpha  \gamma +77\big)\big) r^2\nonumber\\&&\hspace{-0.4cm} +225 \big(6 r^2-5 R^2\big) \alpha ^2+160 \pi ^2 \big(6 r^6 \big(5 \beta ^4-2 \beta ^3+(60 \alpha  \gamma +1) \beta ^2-12 \alpha  \gamma  \beta +2 \alpha  \gamma  (15 \alpha  \gamma +1)\big)-r^4 R^2 \big(30 \beta ^4-13 \beta ^3\nonumber\\&&\hspace{-0.4cm} +(360 \alpha  \gamma +7) \beta ^2-78 \alpha  \gamma  \beta +2 \alpha  \gamma  (90 \alpha  \gamma +7)\big)\big)\big) r^8+4 \rho_s^2 \pi  R^{10} \big(2880 \pi ^2 \gamma  \big(10 \beta ^3-\beta ^2+30 \alpha  \gamma  \beta +\beta -\alpha  \gamma \big) r^6-4 \pi  \big(225 \nonumber\\&&\hspace{-0.4cm} \times\big(32 \pi  R^2 \gamma -9\big) \beta ^3+\big(270-1120 \pi  R^2 \gamma \big) \beta ^2-81 (150 \alpha  \gamma +1) \beta +80 \pi  R^2 \gamma  (270 \alpha  \gamma +11) \beta +20 \alpha  \gamma  \big(27-56 \pi  R^2 \gamma \big)\big) r^4\nonumber\\&&\hspace{-0.4cm} +\big(-4 \pi  \beta  \big(1725 \beta ^2-290 \beta +101\big) R^2-5 \alpha  \big(8280 \pi  \beta  \gamma  R^2-464 \pi  \gamma  R^2-675 \beta +81\big)\big) r^2+135 R^2 \alpha  (3-20 \beta )\big) r^6+2 \rho_s \nonumber\\&&\hspace{-0.4cm} \times R^{12} \big(1920 \pi ^3 \gamma ^2 \big(30 \beta ^2+2 \beta +20 \alpha  \gamma +1\big) r^8-16 \pi ^2 \gamma  \big(160 \pi  \gamma  (15 \alpha  \gamma +1) R^2+75 \beta ^2 \big(48 \pi  R^2 \gamma -23\big)-3 (575 \alpha  \gamma +7)\nonumber\\&&\hspace{-0.4cm} +\beta  \big(90-40 \pi  R^2 \gamma \big)\big) r^6-40 \pi  \big(6 \pi  \gamma  (95 \alpha  \gamma +2) R^2-90 \alpha  \gamma +15 \beta ^2 \big(38 \pi  R^2 \gamma -3\big)+\beta  \big(9-50 \pi  R^2 \gamma \big)\big) r^4-15 \big(2 \pi  \beta  (45 \beta \nonumber\\&&\hspace{-0.4cm} -13) R^2+15 \alpha  \big(12 \pi  R^2 \gamma -1\big)\big) r^2-150 R^2 \alpha \big) r^2+5 R^{14} \big(768 \pi ^3 (10 \beta +1) \gamma ^3 r^8-240 \pi ^2 \beta  \gamma ^2 \big(32 \pi  R^2 \gamma -19\big) r^6-4 \pi  \gamma  \nonumber\\&&\hspace{-0.4cm} \times\big(-40 \pi  \gamma  R^2+45 \beta  \big(20 \pi  R^2 \gamma -3\big)+9\big) r^4+30 \big(-12 \pi  \beta  \gamma  R^2+2 \pi  \gamma  R^2+\beta \big) r^2-15 R^2 \beta \big),\nonumber\\
 &&\hspace{-1.2cm} \mathcal{F}_{10}(r)=15 \big(13 r^4-22 R^2 r^2+9 R^4\big) \alpha +8 \pi  \big(\big(90 \beta ^2-42 \beta +60 \alpha  \gamma +3\big) r^6-R^2 \big(180 \beta ^2-83 \beta +120 \alpha  \gamma +7\big) r^4+R^4 \big(90 \beta ^2\nonumber\\&&\hspace{-0.4cm} -33 \beta +60 \alpha  \gamma +4\big) r^2\big),\nonumber\\
&&\hspace{-1.2cm} \mathcal{F}_{11}(r)=80 \pi  \big(10 \beta ^3-5 \beta ^2+30 \alpha  \gamma  \beta +\beta -5 \alpha  \gamma \big) r^6+\big(\alpha  \big(-4800 \pi  \beta  \gamma  R^2+816 \pi  \gamma  R^2+875 \beta -60\big)-16 \pi  R^2 \beta  \big(100 \beta ^2-51 \beta\nonumber\\&&\hspace{-0.4cm}  +12\big)\big) r^4+2 \big(8 \pi  \big(50 \beta ^3-18 \beta ^2+150 \alpha  \gamma  \beta +7 \beta -18 \alpha  \gamma \big) R^4+\alpha  (59-725 \beta ) R^2\big) r^2+23 R^4 \alpha  (25 \beta -2),\nonumber\\
 &&\hspace{-1.2cm} \mathcal{F}_{12}(r)=\big(480 \pi ^2 \big(r^2-R^2\big) \big(6 \big(5 \beta ^4-2 \beta ^3+(60 \alpha  \gamma +1) \beta ^2-12 \alpha  \gamma  \beta +2 \alpha  \gamma  (15 \alpha  \gamma +1)\big) r^4-3 R^2 \big(20 \beta ^4-9 \beta ^3+5 (48 \alpha  \gamma +1) \beta ^2\nonumber\\&&\hspace{-0.4cm} -54 \alpha  \gamma  \beta +10 \alpha  \gamma  (12 \alpha  \gamma +1)\big) r^2+R^4 \big(30 \beta ^4-7 \beta ^3+9 (40 \alpha  \gamma +1) \beta ^2-42 \alpha  \gamma  \beta +18 \alpha  \gamma  (10 \alpha  \gamma +1)\big)\big) r^4+4 \pi  \alpha  \big(18 \big(775 \beta ^2\nonumber\\&&\hspace{-0.4cm} -90 \beta +775 \alpha  \gamma +11\big) r^6-9 R^2 \big(4050 \beta ^2-545 \beta +4050 \alpha  \gamma +77\big) r^4+6 R^4 \big(5175 \beta ^2-735 \beta +5175 \alpha  \gamma +128\big) r^2-5 R^6\nonumber\\&&\hspace{-0.4cm} \times \big(1710 \beta ^2-225 \beta +1710 \alpha  \gamma +53\big)\big) r^2+2025 \big(r^2-R^2\big)^2 \big(2 r^2-R^2\big) \alpha ^2\big),\nonumber\\
 &&\hspace{-1.2cm} \mathcal{F}_{13}(r)=-2880 \pi ^2 \gamma  \big(10 \beta ^3-\beta ^2+30 \alpha  \gamma  \beta +\beta -\alpha  \gamma \big) r^{10}+12 \pi  \big(225 \big(32 \pi  R^2 \gamma -3\big) \beta ^3+\big(90-1120 \pi  R^2 \gamma \big) \beta ^2-27 (150 \alpha  \gamma +1) \beta \nonumber\\&&\hspace{-0.4cm} +80 \pi  R^2 \gamma  (270 \alpha  \gamma +11) \beta +20 \alpha  \gamma  \big(9-56 \pi  R^2 \gamma \big)\big) r^8-3 \big(4 \pi  \beta  \big(1040 \pi  \gamma  R^2+75 \beta ^2 \big(96 \pi  R^2 \gamma -23\big)+\beta  \big(290-880 \pi  R^2 \gamma \big)\nonumber\\&&\hspace{-0.4cm} -101\big) R^2+5 \alpha  \big(-704 \pi ^2 \gamma ^2 R^4+464 \pi  \gamma  R^2+45 \beta  \big(384 \pi ^2 \gamma ^2 R^4-184 \pi  \gamma  R^2+5\big)-27\big)\big) r^6+5 \big(960 \pi ^2 \beta  \gamma  \big(6 \beta ^2+18 \alpha  \gamma \nonumber\\&&\hspace{-0.4cm} +1\big) R^6-4 \pi  \big(855 \beta ^3-150 \beta ^2+5130 \alpha  \gamma  \beta +71 \beta -300 \alpha  \gamma \big) R^4+81 \alpha  (20 \beta -3) R^2\big) r^4+5 \big(20 \pi  R^6 \big(45 \beta ^3-6 \beta ^2+270 \alpha  \gamma  \beta \nonumber\\&&\hspace{-0.4cm} +5 \beta -12 \alpha  \gamma \big)-9 R^4 \alpha  (135 \beta -23)\big) r^2+225 R^6 \alpha  (6 \beta -1). \nonumber
\end{eqnarray}
\end{small}

\end{document}